# Complex precipitation pathways in multi-component alloys

EMMANUEL CLOUET[1*], LUDOVIC LAÉ[2], THIERRY ÉPICIER[3], WILLIAMS LEFEBVRE[4], MAYLISE NASTAR[1] AND ALEXIS DESCHAMPS[2]

[1] Service de Recherches de Métallurgie Physique, CEA/Saclay, 91191 Gif-sur-Yvette, France.
[2] LTPCM/ENSEEG, UMR CNRS 5614, Domaine Universitaire, BP 75, 38402 St Martin d'Hères, France.
[3] Groupe d'Etudes de Métallurgie Physique et de Physique des Matériaux, UMR CNRS 5510, INSA, 69621 Villeurbanne, France.
[4] Groupe de Physique des Matériaux, UMR CNRS 6634, Université de Rouen, 76801 Saint Étienne du Rouvray, France.
*e-mail: emmanuel.clouet@cea.fr

One usual way to strengthen a metal is to add alloying elements and to control the size and the density of the precipitates obtained. However, precipitation in multicomponent alloys can take complex pathways depending on the relative diffusivity of solute atoms and on the relative driving forces involved. In Al–Zr–Sc alloys, atomic simulations based on first-principle calculations combined with various complementary experimental approaches working at different scales reveal a strongly inhomogeneous structure of the precipitates: owing to the much faster diffusivity of Sc compared with Zr in the solid solution, and to the absence of Zr and Sc diffusion inside the precipitates, the precipitate core is mostly Sc-rich, whereas the external shell is Zr-rich. This explains previous observations of an enhanced nucleation rate in Al–Zr–Sc alloys compared with binary Al-Sc alloys, along with much higher resistance to Ostwald ripening, two features of the utmost importance in the field of light high-strength materials.

Optimizing the properties of an alloy requires a detailed knowledge of its precipitation kinetics. However, the effect of adding distinct impurities to a metal does not reduce to summing the effect of each of them. A spectacular example is the addition of Zr and Sc elements to aluminium alloys. Both elements, when introduced separately, increase the tensile strength and the recrystallization resistance of the alloy by forming ordered precipitates, but the combined effect is considerably larger because it leads to a higher density of smaller precipitates that are also less sensitive to coarsening[1–6]. Although this combined effect has been evaluated in the literature, no satisfactory explanation is available yet. To better understand the precipitation kinetics in Al–Zr–Sc, we present the results of a multiscale approach. Our approach combines atomic-scale simulations, which correctly predict the unusual microstructure of the precipitates in this ternary alloy and the kinetic pathway which leads to it, and experimental techniques, which prove that the simulations reproduce the real material. High-resolution electron microscopy (HREM) and three-dimensional atom probe (3DAP) analysis are used to characterize the atomic-scale distribution of the chemical species in the particles. Small-angle X-ray scattering (SAXS) confirms these results on a larger number of particles, and gives quantitative information on the overall precipitation kinetics. The purpose is to illustrate how the coupling between modelling and experimental techniques at various scales allows the kinetic pathway to be tracked and quantified.

An atomic-diffusion model has previously been developed for both binary Al–Zr and Al–Sc systems[7]. It relies on a rigid lattice with interactions between first- and second-nearest neighbours, and uses a thermally activated atom–vacancy exchange mechanism to describe diffusion. The model has been validated by a mesoscopic extrapolation relying on cluster dynamics showing that experimental data on precipitation are well reproduced[8].

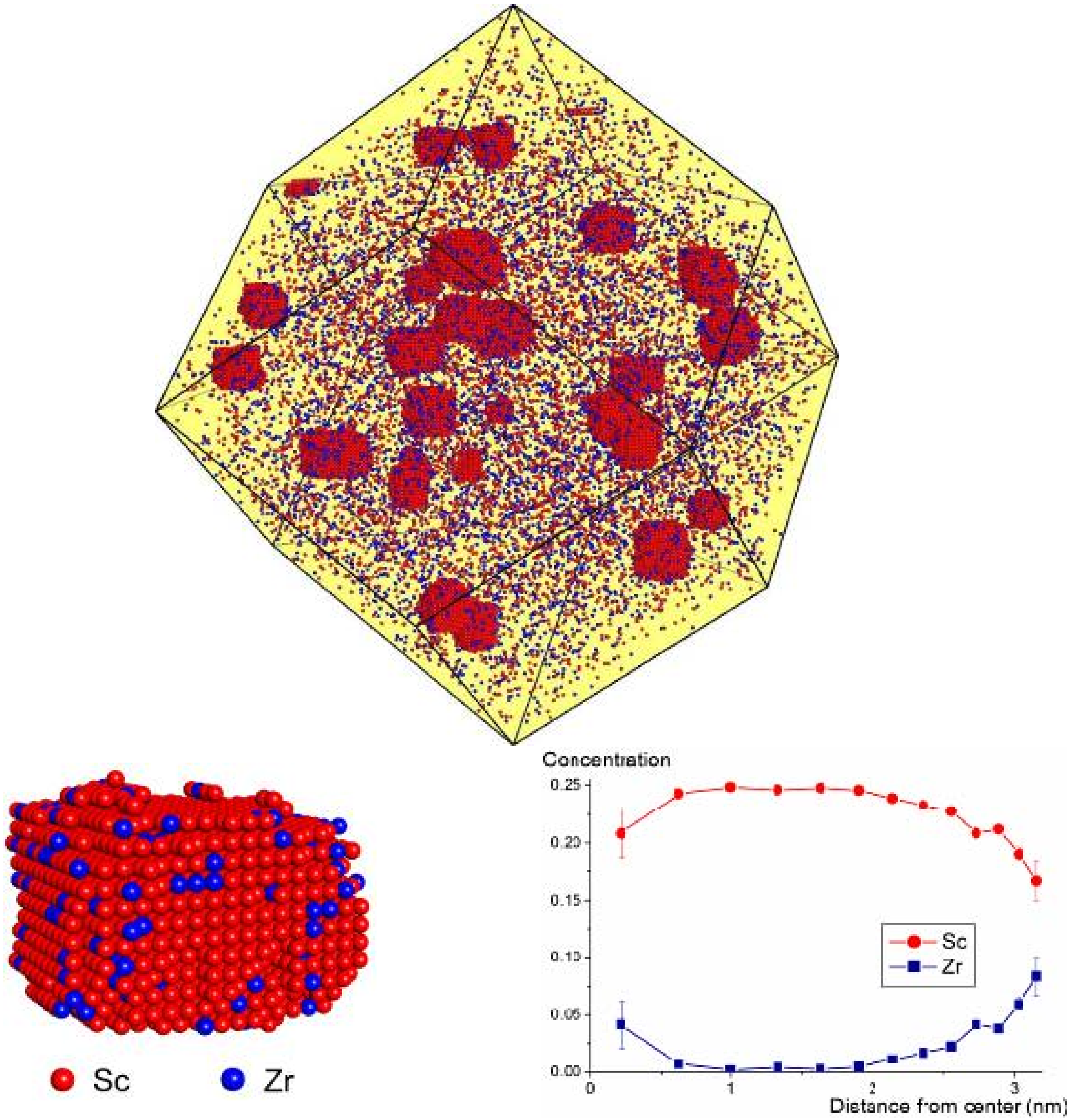

Figure 1 $Al_3Zr_xSc_{1-x}$ **precipitates obtained by atomic simulation.** The positions of Sc and Zr atoms in the computational box are shown after a simulated annealing at 550°C for 1.2 s of an aluminium solid solution containing 0.1 at.% Zr and 0.5 at.% Sc (KMC). A precipitate is isolated and its radial concentration profile drawn, showing the Zr enrichment of the precipitate periphery compared with its core. The error bars correspond to concentration variations associated with the detection of one more or one less solute atom.

To study precipitation in the ternary Al–Zr–Sc system, this atomic model needs to be generalized by including interactions between Zr and Sc atoms. As no experimental information is available, we use a first principle approach to estimate the required parameters. *Ab initio* calculations are carried out to obtain the formation energies of 19 ordered compounds in the Al–Zr–Sc ternary system. Then, we use the inverse Connolly–Williams method[9] to deduce the unknown interactions of the atomic model from this database of formation energies (see the Supplementary Information). We obtain an order energy $\omega^{(1)}_{ZrSc}$ = 237 meV corresponding to a strong repulsion when Sc and Zr atoms are first-nearest neighbours, and a slight attraction when they are second-nearest neighbours ($\omega^{(2)}_{ZrSc}$ = –2.77 meV). With such interactions, an ordered ternary compound $Al_6ZrSc$ is stable at 0 K, however, as soon as the temperature is higher than 70 K, the model predicts that it partially disorders and leads to an $Al_3Zr_xSc_{1-x}$ compound, where $0 \leq x \leq 1$ is a variable quantity. This compound has a $L1_2$ structure, and the atoms on the minority sublattice can equally be Zr or Sc. This agrees with experimental observations[10] carried out on massive $Al_3Zr_xSc_{1-x}$ samples with transmission electron microscopy (TEM), which show that above the ambient temperature precipitates in the ternary system have the structure described above.

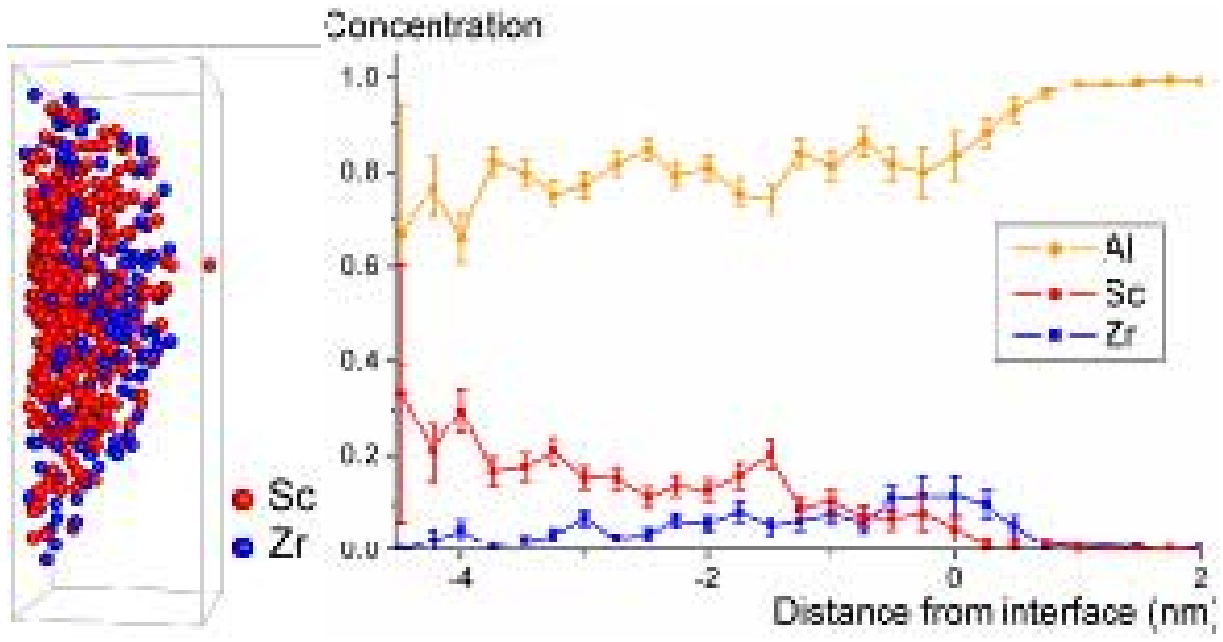

**Figure 2** **3DAP analysis of the edge of an $Al_3Zr_xSc_{1-x}$ precipitate.** A precipitate was isolated in an aluminium solid solution containing 0.09 at.% Sc and 0.03 at.% Zr aged for 128 h at 400°C. The corresponding 5×5×18 $nm^3$ volume is shown on the left, each sphere representing a solute atom. The associated concentration profile on the right shows the Zr enrichment of the precipitate in its periphery. The error bars correspond to concentration variations associated with the detection of one more or one less solute atom.

Using this atomic-diffusion model, we run kinetic Monte Carlo (KMC) simulations of the annealing of supersaturated Al–Zr–Sc solid solutions. $Al_3Zr_xSc_{1-x}$ precipitates appearing in the simulation box are inhomogeneous (Fig. 1): their core is richer in Sc than in Zr. The Zr concentration slightly decreases away from the core, and then strongly increases at the periphery of the precipitates, thus forming a Zr-rich shell. This strong Zr segregation at the interface, as predicted by our model[11], agrees with experimental observations with

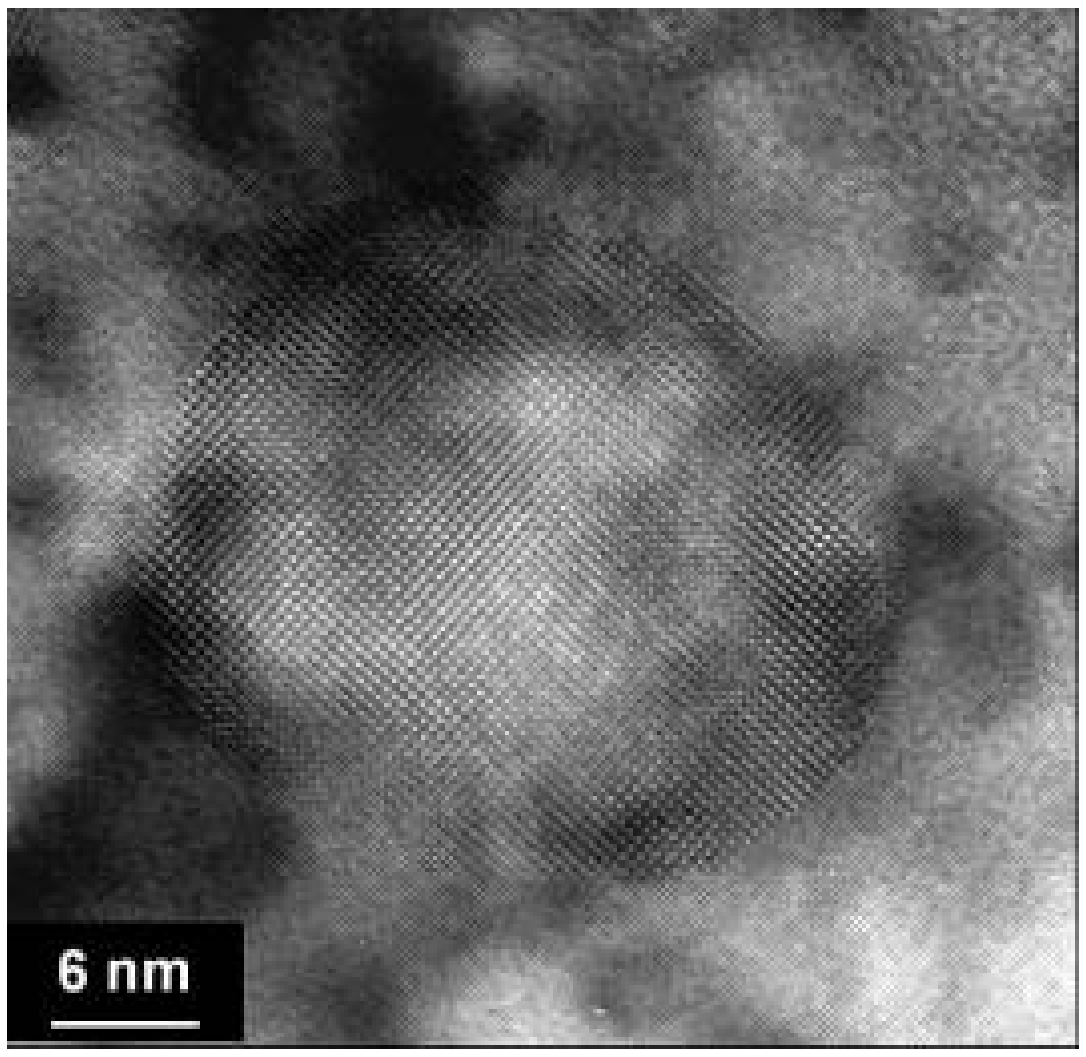

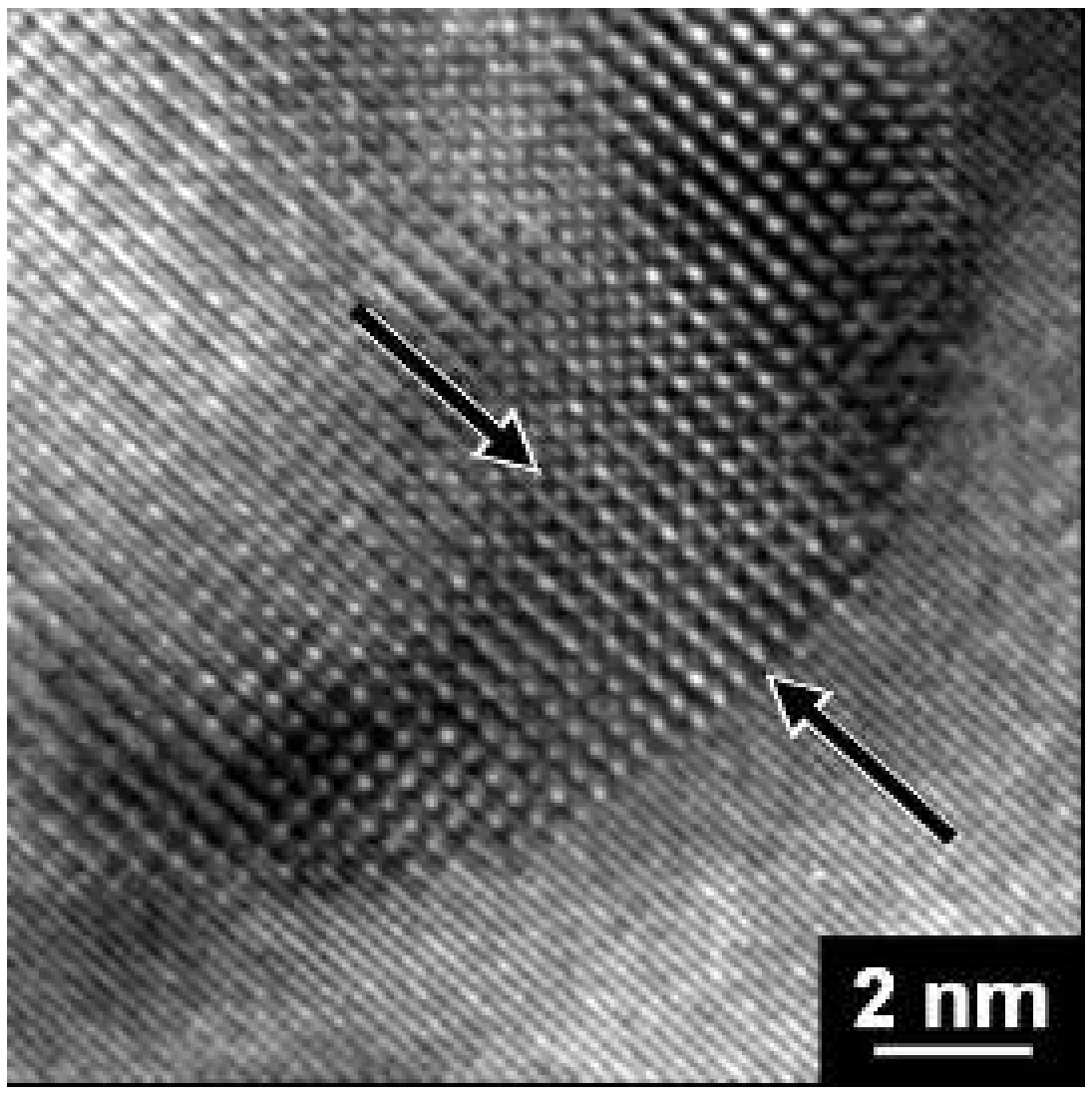

**Figure 3** **HREM image of an $Al_3Zr_xSc_{1-x}$ precipitate.** The particle has been observed in an aluminium solid solution containing 0.09 at.% Sc and 0.03 at.% Zr aged during 32 h at 450°C. The bottom-right side of the whole precipitate shown in the top picture is enlarged in the bottom picture, which shows the higher contrast of the shell region between the arrows, owing to the increase of the heavier Zr atoms.

HREM[12] and with 3DAP[13–15] on similar alloys. We also carried out observations, all on the same model Al–Zr–Sc alloy, to allow a direct comparison between these two different atomic-scale characterization techniques. In addition, SAXS is used to characterize the precipitation process on a larger scale, and thus to obtain statistically relevant information[16]. Alloys corresponding to a supersaturated solid solution containing 0.09 at.% Sc and 0.03 at.% Zr are prepared. To ensure that no microsegregation is present, they are homogenized for 360 h at 630°C. One sample is then aged for 128 h at 400°C and another one for 32 h at 450°C.

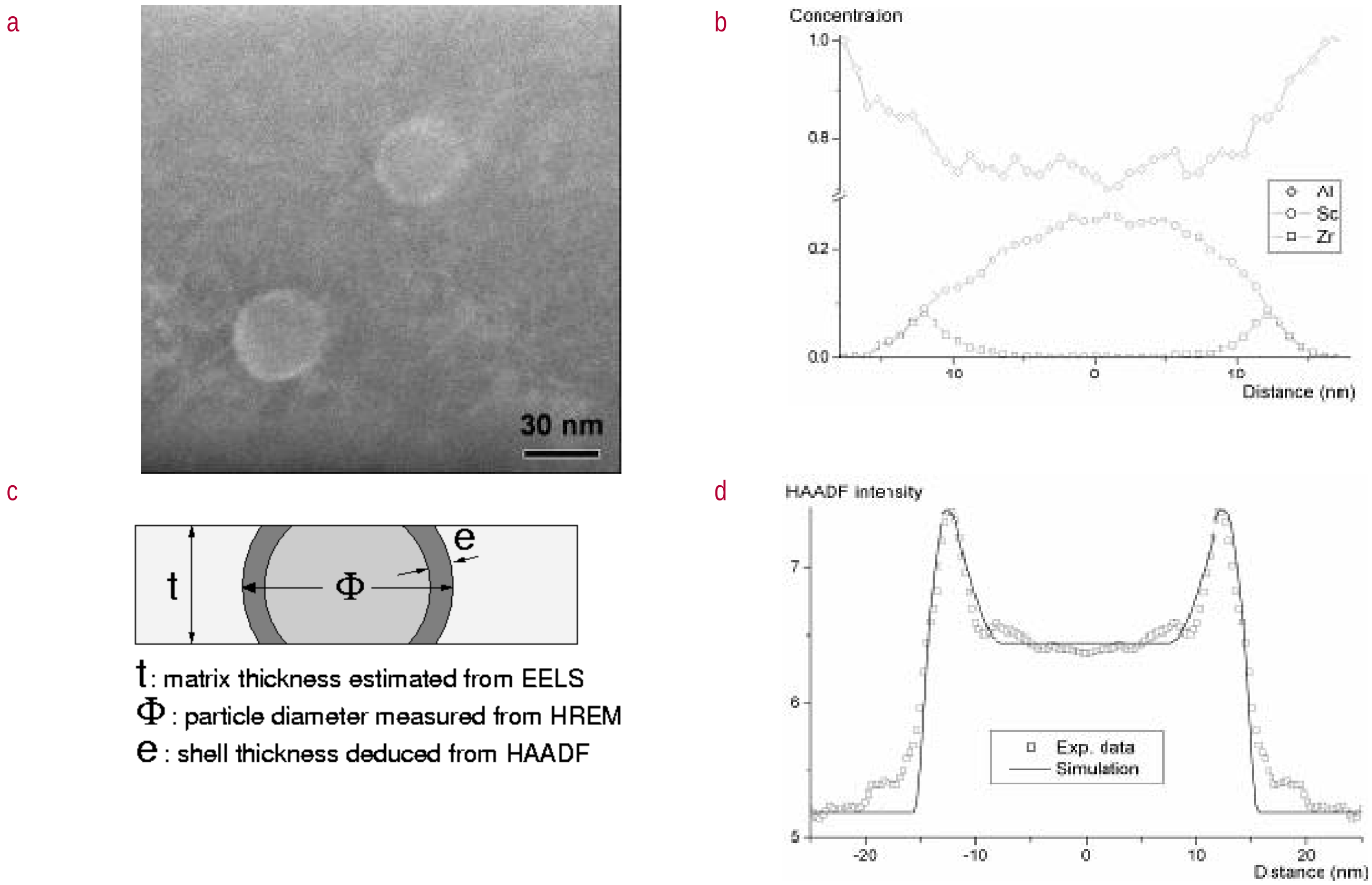

Figure 4 **HAADF analysis of $Al_3Zr_xSc_{1-x}$ precipitates.** a, HAADF image of particles observed in an aluminium solid solution containing 0.09 at.% Sc and 0.03 at.% Zr aged for 32 h at 450°C showing the bright contrast associated with the Zr-enriched shell. b, The EDX line-scan of the bottom particle. c, The geometrical model used to simulate the HAADF contrast. d, The rotationally averaged profile of the HAADF intensity from the bottom particle agrees with the simulated one for precipitates with core and shell compositions equal to $Al_3Sc$ and $Al_3Zr_{0.5}Sc_{0.5}$, respectively.

3DAP analyses reveal particles enriched in Zr and Sc (Fig. 2). A closer look suggests that Zr and Sc are not equally distributed in the particle. The interface seems to be richer in Zr and poorer in Sc than the precipitate core. To quantify concentration fluctuations from the interface to the core of the particle, erosion concentration profiles[17] are calculated (Fig. 2). It should be noted that the position of the particle/matrix interface can only be estimated roughly because of trajectory overlaps involved in the material, thus leading to an uncertainty on the concentration profile obtained. Nevertheless, the Zr enrichment is clearly visible close to the edge of the particle, and the thickness of the Zr-rich shell, wherein Zr concentration reaches a maximum of 15 at.%, can be estimated as 3 nm. Conversely, Sc concentration progressively increases and tends towards 25 at.% as we approach the core of the particle. As for the Al concentration, it fluctuates between 65 and 85 at.%, and is thus compatible with a 75 at.% average value corresponding to the $Al_3Zr_xSc_{1-x}$ stoichiometry.

To confirm the 3DAP observations, HREM, high-angle annular dark-field (HAADF) imaging, energy-dispersive X-ray (EDX) and electron energy-loss spectroscopy (EELS) analyses are carried out. The results are shown in Figs 3 and 4. The representative precipitate of Fig. 3 is observed in HREM along the [001] Al-matrix azimuth. Its diameter measures about 26 nm, and the contrast clearly shows a core–shell configuration with a non-uniform shell thickness of about 2–4 nm as seen from the bottom-right side of the particle. In HAADF imaging (Fig. 4a), the shell appears as an outer ring that is brighter than the core. HAADF images are recorded under conditions excluding any diffraction effect and, in these observation conditions, the image contrast is known to be roughly proportional to the square of the atomic number[18]. As Zr has a higher atomic number than Sc ($^{40}$Zr and $^{21}$Sc), the bright shell corresponds to a Zr-enriched part of the precipitates. An EDX line-scan through the particle located at the bottom of Fig. 4a confirms the Zr-enrichment at the periphery of the particle (Fig. 4b). It indicates a similar Zr and Sc atomic concentration within the shell, that is an $Al_3Zr_{0.5}Sc_{0.5}$ composition, whereas the core tends to have the $Al_3Sc$ stoichiometry. From the estimation of the thickness of the Al-matrix with EELS[19], the HAADF intensity can be simulated on the basis of trial-and-error geometrical modelling of the particle (Fig. 4c). This simulation is based on the thermal diffuse scattering recorded from the different regions of the object, that is, the matrix, the core and the shell of the precipitate, assuming a homogeneous atomic concentration in each of these. The results of these calculations show that a structure composed of a core and a shell with the respective compositions $Al_3Sc$ and $Al_3Zr_{0.5}Sc_{0.5}$ is compatible with the HAADF contrast (Fig. 4d) observed experimentally, and is thus in agreement with the EDX line-scan. The thickness of the shell deduced from this fitting of the HAADF contrast is 2.5 nm. These findings confirm the core–shell structure shown by HREM, and are in qualitative agreement with the results from the 3DAP analysis. All of the characterization techniques lead to similar values for the composition and the thickness of the precipitate external shell.

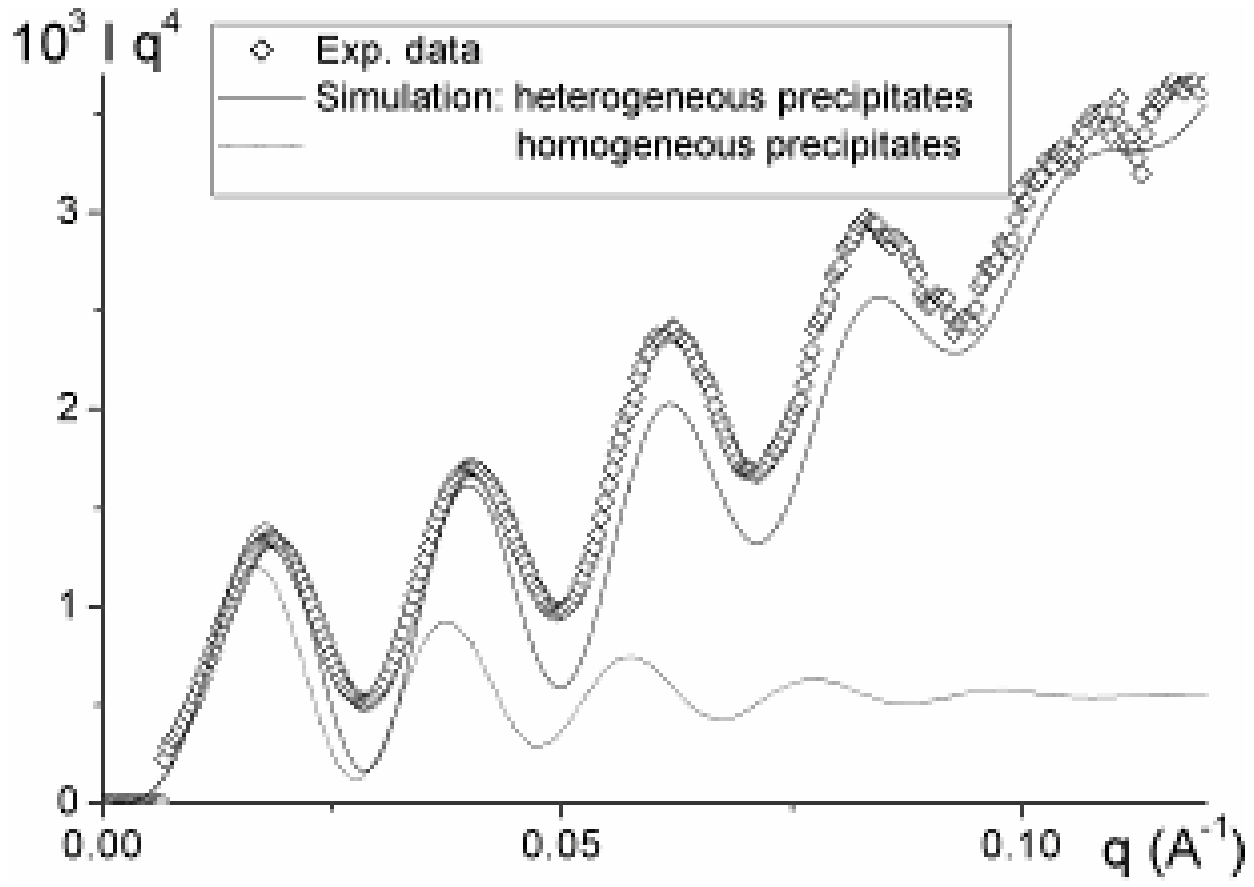

Figure 5 **Small angle X-ray scattering.** The spectrum is measured in an aluminium solid solution containing 0.09 at.% Sc and 0.03 at.% Zr aged for 32 hours at 450°C, and is compared with the signal simulated assuming homogeneous or heterogeneous precipitates.

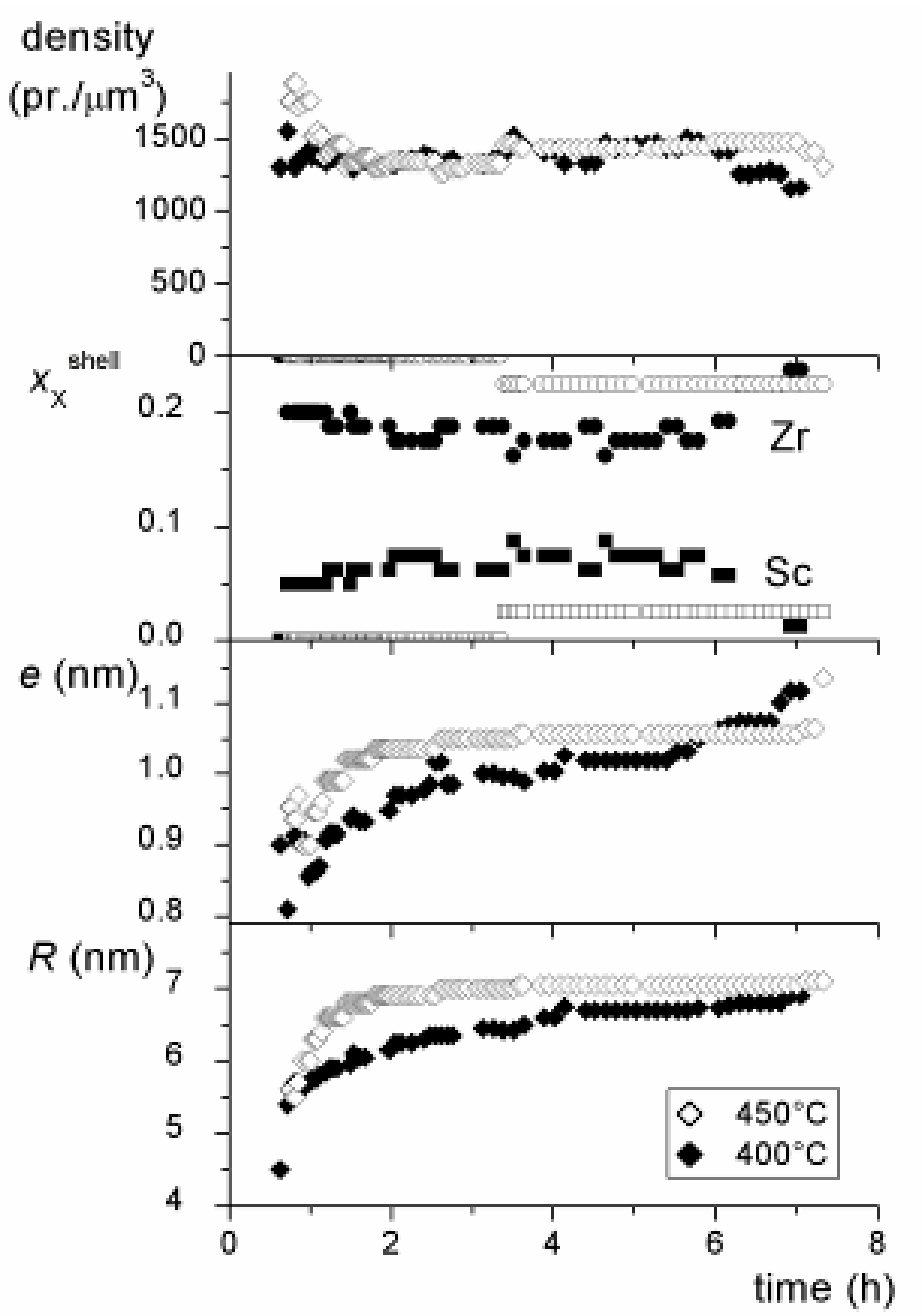

Figure 6 **Precipitation kinetics deduced from SAXS.** The evolution with time of the precipitate radius *R*, of the thickness *e* of the external shell and of its concentrations $x_{Sc}^{shell}$ and $x_{Zr}^{shell}$, and of the precipitate density measured *in situ* for an aluminium solid solution containing 0.09 at.% Sc and 0.03 at.% Zr and aged at two different temperatures.

Despite their powerful ability to observe precipitates, one drawback of 3DAP analysis and HREM is that they are local observation techniques by nature: they fail to give a quantitative characterization of the overall precipitation kinetics. In contrast, SAXS has the ability to study the precipitation kinetics on a large volume of matter (~$10^{-3}$ mm$^3$) and to obtain average quantities, such as the precipitate mean size and the precipitate volume fraction. SAXS spectra[20] realized on alloy samples aged at 400 and 450°C show strong oscillations of the scattered intensity $I(q)$, with the scattering vector $q$ related to the presence of precipitates. The plot showing $I(q)\,q^4$ versus $q$ (Fig. 5) highlights these oscillations as well as an unusual strong linear slope. Such a signal cannot be interpreted as originating from a distribution of chemically homogeneous precipitates. This is shown in Fig. 5 by comparing the experimental scattering spectrum with a simulated one, taking the precipitate-size distribution from TEM image analyses and assuming homogeneous precipitates: the simulated signal is oscillating but does not show any slope contrary to the experimental signal. This linear behaviour actually arises from the contrast of the precipitate electronic density associated with their composition heterogeneity, as seen above with atomic simulations and local characterization techniques. The experimental scattering intensity can be adequately reproduced by a simulation of the SAXS signal assuming precipitates are composed of a pure $Al_3Sc$ core and a concentric shell $Al_3Zr_xSc_{1-x}$ of unknown composition *x*. Fitting the simulated signal to the experimental one precisely yields not only the mean radius, the standard deviation of the size distribution and the volume fraction of the precipitates, but also the composition and the relative thickness of the external shell. In the alloy samples aged for 128 h at 400°C (32 h at 450°C), precipitates have a mean radius equal to 16.5 nm (14 nm); their external shell is 1-nm (2-nm) thick and contains 17.5% Zr and 7.5% Sc (19% Zr and 6%Sc). The thickness of the external shell deduced from SAXS is slightly lower than that observed with 3DAP and HREM, whereas the Zr concentration of the shell is slightly higher. Nevertheless, the agreement is reasonable, and thus shows the ability of SAXS measurements to characterize the precipitate heterogeneity on a large scale.

SAXS *in situ* experiments have been carried out (Fig. 6) to follow the time evolution of the precipitate size, their density and composition. For the lower temperature ($T$=400°C), it can be seen that the precipitate size is increasing concurrently with that of the external shell. This shell becomes richer in Zr than in Sc, and tends to have the $Al_3Zr$ stoichiometry after 7 h. This indicates that precipitates are mainly growing by absorbing zirconium. At the higher temperature ($T$=450°C), precipitates do not grow and the external shell does not evolve further after 2 h: once the shell has reached the $Al_3Zr$ composition, the precipitate size becomes stable. The resistance to coarsening is further shown by the evaluation of the precipitate density, which remains constant with time. These two experiments show that the precipitates formed in the Al–Zr–Sc alloys have a remarkable stability, especially once the concentration of the shell reaches a sufficient value corresponding roughly to the $Al_3Zr$ stoichiometry.

Because of computational time limitations, KMC simulations are processed on higher supersaturations. Nevertheless, they lead to the same qualitative results and allow us to understand the precipitation kinetic path, especially the observed core–shell structure. Different explanations for the Zr segregation at the interface between the precipitates and the solid solution can be put forward. It can be thought of as an equilibrium segregation associated with an elastic or a thermodynamic stabilization of the interface as well as a kinetic segregation due to the formation history of the precipitates. Indeed, Harada and Dunand[10] showed that the substitution by Zr of the Sc atoms in the $Al_3Sc$ compound leads to a decrease in the lattice parameter. Therefore, it is expected that a Zr segregation at the interface between the precipitates and the solid solution is favoured by a decrease in the lattice mismatch, and thus a decrease in the elastic stress which is required to maintain coherency. Note that the atomic model described above relies on a rigid lattice and does not consider such an elastic contribution. The KMC simulations lead to the Zr segregation, showing that part of the heterogeneity of the precipitate composition has an origin other than an elastic one. The second possibility is a thermodynamic segregation due to a lowering of the interface free energy because of chemical interactions between the different atomic species. Marquis *et al.*[21,22] showed that such a thermodynamic effect exists in Al–Sc–Mg alloys: it explains the Mg segregation at the interface between $Al_3Sc$ precipitates and the aluminium solid solution. To check whether this is also the case in Al–Zr–Sc alloys, we run some KMC simulations with modified atomic parameters: we decrease the Sc attempt frequency to slow down the Sc diffusion until the same diffusion coefficients for Sc and Zr atoms are obtained. In doing this, we only change the kinetic behaviour of the solute atoms and do not modify the thermodynamics of the ternary Al–Zr–Sc system. In particular, the interface free energies are the same in both atomic models. The precipitates obtained with these modified atomic parameters are homogeneous and do not show any segregation of the Zr atoms at the interface with the solid solution. This clearly demonstrates that the core–shell structure of the precipitates originates from the difference between the diffusion coefficients of the solute atoms, and that a lowering of the interface free energy cannot be responsible for this observed inhomogeneous structure. Such a lowering may exist, but it is too small to lead to any visible effect on the precipitate structure. To quantify it, we calculate the Zr segregation energy for the three different directions [100], [110] and [111]. A simple count of broken bonds leads to

$$E_{100}^{seg} = \frac{1}{2}E_{110}^{seg} = \frac{1}{3}E_{111}^{seg} = a^2\sigma_{100}(\mathrm{Al_3Zr}) - a^2\sigma_{100}(\mathrm{Al_3Sc}) - \omega_{\mathrm{ZrSc}}^{(2)},$$

where $a$ is the aluminium lattice parameter and $\sigma_{100}(\mathrm{Al_3X})$ is the interface free energy for the [100] direction between the $Al_3X$ precipitate and the aluminium solid solution. As the two binary compounds $Al_3Zr$ and $Al_3Sc$ have similar interface free energies[7], and as the order energy between Sc and Zr atoms is really low in magnitude when they are second-nearest neighbours, the Zr segregation energies for the different orientations of the interface are small: at 500°C, the segregation energy is smaller than 1.5 meV for the three directions of the planar interface, leading to a factor smaller than 1.02 between the Zr concentrations at the interface and in the core of the precipitate in the dilute limit. Therefore, our atomic model does not predict any relevant equilibrium segregation.

To understand the precipitate heterogeneity, it is necessary to look more precisely at the solute kinetic behaviour. Indeed, Sc atoms diffuse faster than Zr atoms: the ratio of the diffusion coefficients $D_{Sc^*}/D_{Zr^*}$ varies from 1,800 to 190 between 400 and 550°C. Therefore, during the first stages of the decomposition of the solid solution, precipitates preferentially grow by absorbing Sc atoms. When the Sc concentration in the solid solution approaches its equilibrium value, precipitates can only grow by absorbing Zr atoms. Because of the high energy cost associated with the creation of anti-site defects in the $Al_3Zr_xSc_{1-x}$ compound, there is no diffusion of the solute atoms inside the precipitates (see the Supplementary Information), which cannot rearrange themselves to reach their equilibrium structure. This means that the core–shell structure is a direct consequence of the formation history: the Sc-rich core corresponds to the early stage of growth and the Zr rich shell corresponds to the late stage when the solid solution is depleted in Sc and precipitates are growing by Zr absorption. Thus, in predicting kinetic pathways in multicomponent alloys, atomic simulations are essential. Indeed, other models of the precipitation working at larger scale, such as cluster dynamics[23] or models based on classical descriptions of the nucleation, growth and coarsening stages[24], assume that precipitates have their equilibrium structure. Therefore, they cannot predict any heterogeneity of the precipitate composition. On the other hand, by taking full account of the vacancy mechanism leading to the decomposition of a supersaturated solid solution, KMC simulations do not require such a strong assumption and naturally lead to the structure of the precipitates, this structure being at equilibrium or not.

The atomic model also allows us to understand why the addition of Zr to an Al–Sc solid solution leads to an increase in the precipitate density and a decrease in their size as observed experimentally[4–6] and reproduced by our KMC simulations (see the Supplementary Information). Indeed, such an addition increases the nucleation driving force because of the configuration entropy associated with the disordered minority sublattice of the $Al_3Zr_xSc_{1-x}$ $L1_2$ structure. This entropic contribution can be seen on the structure of the precipitates (Fig. 1): it explains the higher Zr concentration in the precipitate's inner core formed during the nucleation stage compared with that in the intermediate shells that appeared during the beginning of the growth stage. Unlike the Zr enrichment of the external shells, which has a kinetic origin, this enrichment of the core originates from thermodynamics. This increase in the nucleation driving force results, of course, in an increase in the precipitate density. Thus, the nucleation stage consumes more Sc with this Zr addition than without it, which means that at the end of the nucleation stage there is less Sc available in the solid solution, whereas the number of nuclei is higher. As clusters mainly grow by Sc absorption, precipitates are smaller at the time the solid solution becomes depleted of Sc. Of course, with the Zr addition, precipitates can continue to grow by absorbing Zr atoms, but, because of the difference between the diffusion coefficients of the solute atoms, this second phase of the growth stage is slower than that corresponding to the absorption of Sc atoms. In the ternary alloy, the coarsening stage is controlled by the Zr diffusion: coarsening occurs by evaporating small precipitates to condensate their solute atoms on bigger precipitates, and the Zr-enriched external shell has to be dissolved before a precipitate can be evaporated. This explains the ternary alloy's good resistance to coarsening.

This study illustrates how atomic simulations can be combined with different experimental techniques to obtain a deeper understanding, and hence a better control of the precipitation kinetics. Al–Zr–Sc alloy seems to be one example where the precipitation path is strongly dictated by the diffusion mechanism, and not only by the thermodynamic driving forces.

## METHODS

### KMC

The thermodynamics of the Al–Zr–Sc alloy is described using an Ising model. Thus, atoms are constrained to lie on a face-centred-cubic lattice, and configurations are described by the occupation number $p_n^i$, with $p_n^i = 1$ if the site $n$ is occupied by an atom (or a vacancy) of type $i$ or 0 if not. The energy of a given configuration is

$$E = \frac{1}{2}\sum_{\substack{n,m\\i,j}} \varepsilon_{ij}^{(1)} p_n^i p_m^j + \frac{1}{2}\sum_{\substack{r,s\\i,j}} \varepsilon_{ij}^{(2)} p_r^i p_s^j ,$$

where the first and second sums run on all first- and second-nearest neighbour pairs of sites respectively, and $\varepsilon_{ij}^{(1)}$ and $\varepsilon_{ij}^{(2)}$ are the effective energies of the respective pairs in the configuration $\{i, j\}$. The alloy thermodynamics does not depend on all of the effective energies but only on the order energies $\omega_{ij}^{(n)} = \varepsilon_{ij}^{(n)} - (1/2)\varepsilon_{ii}^{(n)} - (1/2)\varepsilon_{jj}^{(n)}$, where $i$ and $j$ are different species (atoms or vacancy). This model is the simplest one that can be used to model coherent precipitation of $L1_2$ compounds in Al–Zr–Sc alloy (see the Supplementary Information).

Diffusion is described through vacancy jumps. The vacancy exchange frequency with one of its twelve first-nearest neighbours of type $\alpha$ is given by

$$\Gamma_{\alpha-\mathrm{V}} = \nu_\alpha \exp\left(-\frac{E_\alpha^{\mathrm{act}}}{kT}\right)$$

where $\nu_\alpha$ is an attempt frequency, and the activation energy $E_\alpha^{\mathrm{act}}$ is the energy change required to move the $\alpha$ atom from its initial stable position to the saddle-point position. It is computed as the difference between the contribution $e_\alpha^{\mathrm{sp}}$ of the jumping atom to the saddle-point energy and the contribution of the vacancy and of the jumping atom to the initial energy corresponding to the stable position. This last term is obtained by considering all bonds that are broken during the jump. The parameters of the present model have been deduced from experimental data completed by *ab initio* calculations when necessary (see the Supplementary Information).

A time-residence algorithm is used to run KMC simulations. The simulation boxes contain $N_s$ = $100^3$ or $200^3$ lattice sites, and a vacancy occupies one of these sites. At each time step, the vacancy can exchange with one of its twelve first-nearest neighbours, the probability for each of these events being proportional to $\Gamma_{\alpha-\mathrm{V}}$. The time increment corresponding to the current configuration is

$$\Delta t = \frac{1}{N_s(\mathrm{Al})C_\mathrm{V}(\mathrm{Al})}\frac{1}{\sum_{\alpha=1}^{12}\Gamma_{\alpha-\mathrm{V}}},$$

where $C_\mathrm{V}(\mathrm{Al})$ is the real vacancy concentration in pure Al as deduced from energy parameters, and $N_s(\mathrm{Al})$ is the number of lattice sites that can be considered as pure Al (no solute atoms as first- or second-nearest neighbours). The timescale of the simulation is obtained by summing all time increments corresponding to configurations where the vacancy is in pure Al.

## 3DAP

3DAP analyses are carried out on an energy-compensated optical tomographic atom probe (ECOTAP)[25,26]. The edge of a particle is isolated by means of field ion microscopy before 3DAP analyses, the analysis direction being close to a $\langle 114\rangle$ direction, and the observed volume is then reconstructed using a new procedure described elsewhere[27].

## TEM

HREM, HAADF imaging, EDX and EELS analyses are carried out on a JEOL 2010F instrument. Thin foils are prepared by the conventional ion-beam thinning method, and for HAADF and EDX, a subnanometric probe of 0.8 nm is used.

## SAXS

SAXS measurements were all carried out at the D2AM beam line of the European synchrotron radiation facility in Grenoble. A monochromatic beam was used ($\lambda$=1.61 Å). Samples are about 80-$\mu$m thick, and were prepared by mechanical grinding followed by electropolishing. The scattered intensity was measured by a charge-coupled device camera located approximately 1 m from the sample, resulting in a measuring range of scattering vectors $6.8\times10^{-3}$ – 0.2 $\text{Å}^{-1}$. *In situ* experiments are carried out under the X-ray beam with a resistance furnace at a heating rate of 10°C.$\mathrm{min}^{-1}$.

Quantitative information on the precipitate distributions are deduced from SAXS data using the following description of a precipitate of radius *R* with an enriched external shell of thickness $e$. The mean electronic density is taken as equal to $\rho_{\mathrm{core}}$ in the precipitate core (for radii $r < R-e$) and to $\rho_{\mathrm{shell}}$ in the shell (for $R-e<r<R$). $\rho_{\mathrm{core}}$ is calculated assuming an $Al_3Sc$ stoichiometry, whereas $\rho_{\mathrm{shell}}$ is calculated as a function of the composition of the external shell. The intensity scattered in the direction $q$ by a precipitate is

$$I(q,R,e) = \left[F(q,R-e,\rho_{\mathrm{core}}-\rho_{\mathrm{shell}}) + F(q,R,\rho_{\mathrm{shell}})\right]^2,$$

where the scattered amplitude $F$ takes its usual form for homogeneous spherical precipitates:

$$F(q,R,\rho) = \frac{4\pi(\rho-\rho_\mathrm{m})}{q^3}\left(\sin(qR) - qR\cos(qR)\right),$$

$\rho_\mathrm{m}$ being the electronic density of the matrix. We then suppose, in agreement with TEM observations, that the particle size distribution obeys a log-normal law $f(R)$, in which the mean value is $\overline{R}$. Another assumption of the model is that all precipitates have external shells with the same thickness and the same composition. The total scattered intensity is thus given by

$$\mathrm{I}(q) = \int_0^\infty f(R)I(q,R,e)dR .$$

The fitting of this theoretical scattered intensity to the measured experimental signal allows us to obtain the mean radius $\overline{R}$ and the volume fraction of the precipitates as well as the composition and the thickness $e$ of their enriched shell.

## References

1. Yelagin, V. I., Zakharov, V. V., Pavlenko, S. G. & Rostova, T. D. Influence of zirconium additions on ageing of Al-Sc alloys. *Phys. Met. Metall.* **60**, 88-92 (1985).
2. Davydov, V. G., Yelagin, V. I., Sakharov, V. V. & Rostova, T. D. Alloying aluminum alloys with scandium and zirconium additives. *Metal Science and Heat Treatment* **38**, 347-352 (1996).
3. Toropova, L. S., Eskin, D. G., Kharaterova, M. L. & Bobatkina, T. V. *Advanced Aluminum Alloys Containing Scandium – Structure and Properties* (Gordon and Breach Sciences, Amsterdam, 1998).
4. Fuller, C. B., Seidman, D. N. & Dunand, D. C. Mechanical properties of Al(Sc,Zr) alloys at ambient and elevated temperatures. *Acta Mater.* **51**, 4803-4814 (2003).
5. Riddle, Y. W. & Sanders, T. H. A study of coarsening, recrystallization and morphology of microstructure in Al-Sc-(Zr)-(Mg) alloys. *Metal. Mater. Trans. A* **35**, 341-350 (2004).
6. Fuller, C. B. & Seidman, D. N. Temporal evolution of the nanostructure of Al(Sc,Zr) alloys: Part II – Coarsening of $Al_3(Sc_{1-x}Zr_x)$ precipitates. *Acta Mater.* **53**, 5415-5428 (2005).
7. Clouet, E., Nastar, M. & Sigli, C. Nucleation of $Al_3Zr$ and $Al_3Sc$ in aluminum alloys: from kinetic Monte Carlo simulations to classical theory. *Phys. Rev. B* **69**, 064109 (2004).
8. Clouet, E., Barbu, A., Laé, L. & Martin, G. Precipitation kinetics of $Al_3Zr$ and $Al_3Sc$ in aluminum alloys modeled with cluster dynamics. *Acta Mater.* **53**, 2313-2325 (2005).
9. Connolly, J. W. & Williams, A. R. Density-functional theory applied to phase transformations in transition-metal alloys. *Phys. Rev. B* **27**, 5169-5172 (1983).
10. Harada, Y. & Dunand, D. C. Microstructure of $Al_3Sc$ with ternary transition-metal additions. *Mater. Sci. Eng.* **A329–331**, 686-695 (2002).
11. Clouet, E. *Séparation de phase dans les alliages Al-Zr-Sc: du saut des atomes à la croissance de précipités ordonnés* (PhD thesis, École Centrale Paris, 2004) ; http://tel.ccsd.cnrs.fr/tel-00005967.
12. Tolley, A., Radmilovic, V. & Dahmen, U. Segregation in $Al_3(Sc,Zr)$ precipitates in Al-Sc-Zr alloys. *Scripta Mater.* **52**, 621-625 (2005).
13. Fuller, C. B. *Temporal evolution of the microstructures of Al(Sc,Zr) alloys and their influences on mechanical properties*. (PhD thesis, Northwestern University, 2003); http://arc.nucapt.northwestern.edu/refbase/show.php?record=147.
14. Forbord, B., Lefebvre, W., Danoix, F., Hallem, H. & Marthinsen, K. Three dimensional atom probe investigation of the formation of $Al_3(Sc,Zr)$-dispersoids in aluminium alloys. *Scripta Mater.* **51**, 333-337 (2004).
15. Fuller, C. B., Murray, J. L. & Seidman, D. N. Temporal evolution of the nanostructure of Al(Sc,Zr) alloys: Part I – Chemical compositions of $Al_3(Sc_{1-x}Zr_x)$ precipitates. *Acta Mater.* **53**, 5401-5413 (2005).
16. Laé, L. *Étude de la precipitation en dynamique d'amas dans les alliages d'aluminium et dans les aciers.* (PhD thesis, INPG, 2004) ; http://tel.ccsd.cnrs.fr/tel-00009313.
17. Dumont, M., Lefebvre, W., Doisneau-Cottignies, B. & Deschamps, A. Characterisation of the composition and volume fraction of η' and η precipitates in an Al-Zn-Mg alloy by a combination of atom probe, small-angle X-ray scattering and transmission electron microscopy. *Acta Mater.* **53**, 2881-2892 (2005).
18. Crewe, A. V., Langmore, J. P. & Isaacson, M. S. Resolution and contrast in the scanning transmission microscope. *Physical Aspects of Electron Microscopy and*

*Microbeam Analysis* (eds. Siegel, B.M. & Beaman, D. R.), 47 (Wiley, New York, 1975).
19. Egerton, R.F. *Electron Energy Loss Spectroscopy in the Electron Microscope* (Plenum Press, New York, 1996).
20. Glatter, O. & Kratky, O. *Small Angle X-Ray Scattering* (Academic Press, New York, 1982).
21. Marquis, E. A., Seidman, D. N., Asta, M., Woodward, C. & Ozoliņš, V. Mg segregation at Al/$Al_3Sc$ heterophase interfaces on an atomic scale: experiments and computations. *Phys. Rev. Lett.* **91**, 036101 (2003).
22. Marquis, E. A., Seidman, D. N., Asta, M. & Woodward, C. Composition evolution of nanoscale $Al_3Sc$ precipitates in an Al-Mg-Sc alloy: Experiments and computations. *Acta Mater.* **54**, 119-130 (2006).
23. Laé, L. & Guyot, P. Cluster dynamics in Al(Zr,Sc) Alloys. *Proc. 2nd International Conference on Multiscale Materials Modeling* (ed. Ghoniem, N. M.), 272-274 (UCLA Press, Los Angeles, 2004).
24. Robson J. D. A new model for prediction of dispersoid precipitation in aluminium alloys containing zirconium and scandium. *Acta Mater.* **52**, 1409-1421 (2004).
25. Blavette, D., Bostel, A., Sarrau, J. M., Deconihout, B. & Menand, A. An atom probe for three-dimensional tomography. *Nature* **363**, 432-435 (1993).
26. Bémont, E., Bostel, A., Bouet, M., Da Costa, G., Chambreland, S., Deconihout, B. & Hono, K. Effects of incidence angles of ions on the mass resolution of an energy compensated 3D atom probe *Ultramicroscopy* **95**, 231-238 (2003).
27. De Geuser, F., Lefebvre, W., Danoix, F., Vurpillot, F., Forbord, B. & Blavette, D. An improved reconstruction procedure for the correction of local magnification effects in 3DAP. *Surf. and Interf. Analysis* (in press).

## Acknowledgements

The authors are grateful to G. Martin for his invaluable help and advices all along this work and for his careful reading of the manuscript. They thank too M. Athènes, D. Blavette, M. Guttmann, P. Guyot, B. Legrand, D. Seidman, C. Sigli and F. Soisson for fruitful discussions. They are indebted to C. Sigli and to Alcan for providing the heat-treated alloy samples, and to D. Seidman for sending preprints of Ref. 6 and 15 as far back as October 2002. They thank E. Adam for the use of his atomic visualization tool. This work was supported by the joint research program “Precipitation” between Alcan, Arcelor, CNRS, and CEA. E.C. and L.L. acknowledge funding from Alcan. Correspondence and request for materials should be addressed to E.C.

## Competing financial interest

The authors declare they have no competing financial interest.

# Supplementary information

# Complex precipitation pathways in multi-component alloys

EMMANUEL CLOUET[1*], LUDOVIC LAÉ[2], THIERRY ÉPICIER[3], WILLIAMS LEFEBVRE[4], MAYLISE NASTAR[1] AND ALEXIS DESCHAMPS[2]

[1] Service de Recherches de Métallurgie Physique, CEA/Saclay, 91191 Gif-sur-Yvette, France.
[2] LTPCM/ENSEEG, UMR CNRS 5614, Domaine Universitaire, BP 75, 38402 St Martin d'Hères, France.
[3] Groupe d'Etudes de Métallurgie Physique et de Physique des Matériaux, UMR CNRS 5510, INSA, 69621 Villeurbanne, France.
[4] Groupe de Physique des Matériaux, UMR CNRS 6634, Université de Rouen, 76801 Saint Étienne du Rouvray, France.
*e-mail: emmanuel.clouet@cea.fr

We describe in this appendix our motivations for the choice of the atomic diffusion model, i.e. an Ising model with pair interactions between first and second nearest neighbours only. Some details are then given to the way the interactions between Sc and Zr atoms are obtained from first principle calculations and additional KMC simulations are presented so as to complete the results given in the manuscript. We claimed that the heterogeneous structure of the precipitates highlighted by these simulations arises from the absence of solute diffusion inside the precipitates. We therefore used the atomic model so as to demonstrate that such diffusion cannot happen. Finally, we discuss our choice for the interaction energy between Sc atoms and vacancies in aluminium considering recent first principle calculations.

## Atomic model

The atomic diffusion model used in the present work to study precipitation in Al-Zr-Sc alloys relies on that developed in Ref. 1 for binary Al-Zr and Al-Sc alloys. It is based on a rigid lattice description with pair interactions between first and second nearest neighbours. This is the simplest model that can be used to simulate precipitation of a $Al_3Zr_xSc_{1-x}$ compound having the $L1_2$ structure. Indeed, with interactions between first nearest neighbours only, one cannot prevent the majority sub-lattice of the equilibrium $L1_2$ compound to be occupied by solute Sc and Zr atoms[2]. To do so, one needs to consider interactions between at least first and second nearest neighbours. On the other hand, there is no use for Al-Zr-Sc alloys to consider interactions beyond second nearest neighbours as these interactions are significantly lower. This can be shown thanks to a cluster expansion of ab-initio calculations. This was done in Ref. 3 for Al-Zr alloy and in the appendix A of Ref. 4 for Al-Sc alloy. Order energies $\omega_{ij}^{(n)} = \varepsilon_{ij}^{(n)} - (1/2)\varepsilon_{ii}^{(n)} - (1/2)\varepsilon_{jj}^{(n)}$ corresponding to these expansions are displayed on Fig. S1. One clearly sees that, beyond second nearest neighbours, interactions are significantly lower.

We could have also considered interactions for clusters other than pairs. This has been done in the chapter 5 of Ref. 4 where interactions for the tetrahedron and the triangle of first nearest neighbours have been included in the Ising model in addition to pair interactions for first and second nearest neighbours. It was shown that, for ordering binary solid solutions like Al-Zr and Al-Sc alloys, these many-body interactions do not modify the kinetics of precipitation. In particular, the Onsager coefficients defining diffusion in the solid solution are unchanged whether such interactions are included or not, as well as the nucleation free energies and the interface free energies between the precipitates and the solid solution. In the purpose of building an atomic model as simple as possible, we thus choose to neglect these interactions in the present work and keep only pair interactions between first and second nearest neighbours.

As described in Ref. 1, the first nearest-neighbour order energies between Al and Zr and between Al and Sc atoms are directly deduced from the formation free energies of $Al_3Zr$ and $Al_3Sc$ $L1_2$ compounds. These formation free energies are given by ab-initio calculations and they include a vibrational contribution. The second nearest-neighbour order energies are chosen so as to reproduce Zr and Sc solubility limits in aluminium using their low-temperature expansion,

$$x_{\mathrm{X}}^{\mathrm{eq}} = \exp\left(-6\omega_{\mathrm{Al\text{-}X}}^{(2)}/kT\right) + 6\exp\left(-10\omega_{\mathrm{Al\text{-}X}}^{(2)}/kT\right).$$

For Al-Zr interactions, as we want to model precipitation of the metastable $L1_2$ structure of $Al_3Zr$ compound, we use the metastable solubility limits that we previously obtained from ab-initio calculations[3]. For Al-Sc, we use the experimental solubility limit as the $L1_2$ structure of $Al_3Sc$ is stable.

The obtained interactions slightly differ from the ones given by the cluster expansions (Fig. S1). Nevertheless, both sets of parameters lead to the same thermodynamic behaviour for Al-Zr and for Al-Sc alloys. An attraction is observed between Al and solute atoms (Zr or Sc) when they are first nearest neighbours and a repulsion when they are second nearest neighbours. Our fitting procedure ensures that the predictions of the atomic simulations will be as quantitative as possible as the Zr and Sc solubility limits in aluminium corresponding to the atomic model are the correct ones. Moreover interface free energies between $Al_3Zr$ or $Al_3Sc$ precipitates and the solid solution deduced from the present atomic model[1,4] have been shown to lead to correct predictions of the precipitation kinetics. Indeed, using only these interface free energies and the solute diffusion coefficients as input parameters, cluster dynamics manages to reproduce experimental data like the time evolution of the precipitate size distributions[5] and of the electrical resistivity[6] for different annealing temperatures.

With these atomic parameters, the stable phase of the $Al_3X$ compounds (X ≡ Zr or Sc) is the $L1_2$ structure. Indeed, the formation free energies of the different possible structures of this compound are given by

$$\Delta F(\mathrm{Al_3X,\ L1_2}) = 3\omega_{\mathrm{Al\text{-}X}}^{(1)},$$

$$\Delta F(\mathrm{Al_3X,\ DO_{22}}) = 3\omega_{\mathrm{Al\text{-}X}}^{(1)} + \frac{1}{2}\omega_{\mathrm{Al\text{-}X}}^{(2)},$$

$$\Delta F(\mathrm{Al_3X,\ DO_{23}}) = 3\omega_{\mathrm{Al\text{-}X}}^{(1)} + \frac{1}{4}\omega_{\mathrm{Al\text{-}X}}^{(2)}.$$

As the second nearest neighbour interaction between Al and solute atoms is repulsive ($\omega_{\mathrm{Al\text{-}X}}^{(2)} > 0$ for X ≡ Zr or Sc), the $DO_{22}$ and $DO_{23}$ structure have a higher energy than the $L1_2$ structure. Experimentally, the stable structure of $Al_3Sc$ precipitates is $L1_2$ and only this structure has been observed. On the other hand, $Al_3Zr$ has the stable $DO_{23}$ structure, but for small supersaturations of the solid solution, $Al_3Zr$ precipitates with the metastable $L1_2$ structure and precipitates with the $DO_{23}$ structure only appears for prolonged heat treatment and high enough supersaturations. Our atomic model appears thus well-suited to study the first stages of the precipitation kinetics where precipitates can only have the $L1_2$ structure.

## Al-Zr-Sc thermodynamics

The only parameters missing to our model to simulate precipitation in Al-Zr-Sc alloys are the pair energies between Zr and Sc atoms when they are first and second nearest neighbours. As no quantitative experimental information is available, we use an ab-initio approach to deduce these interactions. We calculate with the full potential linear muffin tin orbitals[7] (FP-LMTO) in the generalized gradient approximation[8] (GGA) the energies of several ordered compounds and then use this database to fit the two missing parameters of the atomic model. Details of ab-initio calculations are similar to those given in Ref. 3. A particular care has been taken to obtain the second nearest neighbour order energy $\omega_{\mathrm{Zr\text{-}Sc}}^{(2)}$. Indeed, as we will show below, we find a repulsive interaction between Zr and Sc atoms when they are first nearest neighbours. Low temperature expansions[4,6] show that, in this case, thermodynamics only depend on the second nearest neighbour order energy $\omega_{\mathrm{Zr\text{-}Sc}}^{(2)}$. Consequently, as soon as one manages to show that $\omega_{\mathrm{Zr\text{-}Sc}}^{(1)} > 0$, only the parameter $\omega_{\mathrm{Zr\text{-}Sc}}^{(2)}$ needs to be calculated accurately.

Ten $Al_xZr_ySc_z$ ordered compounds are used to obtain the $\varepsilon_{\mathrm{Zr\text{-}Sc}}^{(1)}$ pair interaction. Their ab-initio energies are calculated for a volume equal to that of Al at equilibrium without any relaxation. The corresponding structures are displayed in Fig. S2. They have been chosen so as to put aside the interaction $\varepsilon_{\mathrm{Zr\text{-}Sc}}^{(2)}$, thus allowing to fit the Ising model with only $\varepsilon^{(1)}_{\text{Zr-Sc}}$ as an unknown parameter. This leads to the effective interaction $\varepsilon^{(1)}_{\text{Zr-Sc}} = -611\,\text{meV}$ corresponding to an order energy $\omega^{(1)}_{\text{Zr-Sc}} = 238\,\text{meV}$. The associated order energy being largely positive, we thus obtain a strong repulsion between Zr and Sc atoms when they are first nearest neighbours. Taking into account this interaction allows to reasonably reproduce the ab-initio formation energies (Tab. S1). The standard deviation between ab-initio calculations and predictions of the Ising model can appear quite large ($\delta E$ = 30 meV/at.). Nevertheless, as already pointed, we only need to show that the first nearest neighbour interaction between Sc and Zr atoms is repulsive as all thermodynamics will then only depend on the second nearest neighbour interaction.

Another set of nine ordered compounds is used to fit the $\varepsilon^{(2)}_{\text{Zr-Sc}}$ pair interaction. All these structures have the stoichiometry $Al_3Zr_xSc_{1-x}$ and have been built from the $L1_2$ structure: only Al atoms are lying on the majority sub-lattice whereas Zr and Sc atoms occupy the minority sub-lattice (Fig. S3). The energies of these $Al_3Zr_xSc_{1-x}$ compounds can be written as a weighted sum of the energies of the $L1_2$ structures of $Al_3Zr$ and $Al_3Sc$ plus a correction arising from the second nearest neighbour pair interaction between Zr and Sc atoms. Therefore, in the Ising model, the energies of these structures do not depend on the parameter $\varepsilon^{(1)}_{\text{Zr-Sc}}$. These energies have been calculated ab-initio, relaxing the volume but keeping fixed all other degrees of freedom (shape of the unit cell and atomic positions). All ternary compounds are found to have a slightly lower energy than the weighted energies of $Al_3Zr$ and $Al_3Sc$ binary compounds (Tab. S2). This stabilization of $Al_3Zr_xSc_1$ ternary compounds has been already shown by Xu and Freeman[9] who calculated the formation energy of $Al_6ZrSc$ $L1_2$-$c$ structure (Fig. S3c). The energy they obtained ($\Delta E = -31$ meV/at.) is lower than the one we calculated ($\Delta E = -4$ meV/at.). This should arise from the atomic sphere approximation (ASA) they used in their LMTO calculations. So as to obtain quantitative results, full potentials calculations like FP-LMTO are required. The fit of the Ising model to this energy database leads to the interaction $\varepsilon^{(2)}_{\text{Zr-Sc}} = -2.77\,\text{meV}$ and thus to the order energy $\omega^{(2)}_{\text{Zr-Sc}} = -2.77$ due to the choice $\varepsilon^{(2)}_{\text{Zr-Zr}} = \varepsilon^{(2)}_{\text{Sc-Sc}} = 0$. This small negative order energy indicates that a Zr-Sc pair of second nearest neighbour is just slightly more stable than Zr-Zr and Sc-Sc pairs. Nevertheless, considering a non-zero $\omega^{(2)}_{\text{Zr-Sc}}$ parameter leads to a better agreement between ab-initio calculations and predictions of the Ising model (Tab. S2). No doubt that the agreement could be better if one included in the Ising models interactions for other clusters than pairs and going beyond second nearest neighbours. Nevertheless, the corresponding effective energies are even smaller than $\varepsilon^{(2)}_{\text{Zr-Sc}}$ and does not lead to any significant change of Al-Zr-Sc thermodynamics: the standard deviation between predictions of the Ising model and ab-initio calculations ($\delta E$ = 2 meV/at.) is already very small. We then prefer to keep the atomic model as simple as we can as long as it has the ability to reproduce experimental observations and to rationalize them. The different parameters of the atomic model are summarized in tables S3 and S4.

## Atomic simulations

As shown in the manuscript, kinetic Monte Carlo simulations lead to the formation of inhomogeneous $Al_3Zr_xSc_{1-x}$ precipitates: their periphery has a higher Zr concentration, thus forming a Zr-enriched shell. This inhomogeneity has been observed for all annealing temperatures and all supersaturations that have been studied. In Fig. 1 in the manuscript, we give the example of a precipitate observed in a solid solution containing 0.1 at.% Zr and 0.5 at.% Sc annealed at 550°C. Some more examples are given in Fig. S4 for a solid solution containing 0.5 at.% Zr and 0.5 at.% Sc annealed at 450°C and 550°C. This thus illustrates more thoroughly the precipitate inhomogeneous structure arising from the difference between the solute diffusion coefficients and from the fact that precipitates cannot rearrange themselves to reach their equilibrium structure.

In agreement with TEM observations[10-12], kinetic Monte Carlo simulations show that a Zr addition to an Al-Sc alloy leads to a higher density of precipitates, these precipitates being smaller (Fig. S5). Indeed, such an addition increases the nucleation driving force and thus the number of precipitates. As these ones mainly grow by absorbing Sc, one obtains smaller precipitates because the number of growing clusters is higher and there is less Sc available for growth at the end of the nucleation stage. The precipitates growth by Zr absorption is really slow but plays an important role too as it leads to the formation of an external Zr enriched shell which is responsible for the good resistance against coarsening. This increase of the nucleation driving force with a Zr addition is responsible for the higher Zr concentration in the inner core of the precipitates (Fig. 1 of the manuscript and Fig. S4a). This inner core forms during the nucleation stage and, as there is no diffusion inside the precipitates, it keeps its composition during the whole kinetics.

## Diffusion inside the precipitates

The heterogeneous structure of the precipitates arises from the absence of Zr and Sc diffusion inside the precipitates. As a consequence, the latter cannot reach their equilibrium structure. In order to support this statement, we deduced from our atomic model estimations of the activation energies involved in Zr and Sc intra-diffusion. At low enough temperatures where $L1_2$ precipitates remain ordered, the minority elements diffuse via highly correlated vacancy sequences, the so-called six-jump vacancy cycles[13-15]. Two different cases have to be considered, depending on which of the two sub-lattices the vacancy is located: vacancy $V_\alpha$ on the majority or $V_\beta$ on the minority sub-lattice of the $L1_2$ structure. Moreover, for each configuration, the vacancy can realize two different cycles, a direct one or a bent one (Fig. S6). Using the atomic diffusion model depicted in the manuscript, we calculate the activation barriers of the jumps occurring in these different cycles for the case of a diffusing impurity Y substituting an X atom of an $Al_3X$ $L1_2$ compound (Tab. S5, S6, S7 and S8). This can be then used to study the diffusion of Sc impurity in $Al_3Zr$ (Fig. S7) and of Zr impurity in $Al_3Sc$ (Fig. S8). In both cases and for both kinds of vacancies ($V_\alpha$ or $V_\beta$), the migration barrier associated with the direct cycle is lower than the one corresponding to the bent cycle. This migration barrier corresponds to the energy difference between the higher saddle point position and the lower stable position, i.e. between the saddle point position corresponding to the jumps from the configuration 2 to 3 and the stable position corresponding to the configuration 0. This migration energy is thus given by

$$E_{\mathrm{Y}}^{\mathrm{mig}}(\mathrm{Al_3X}) = E_{0\to1}^{\mathrm{mig}} - E_{1\to0}^{\mathrm{mig}} + E_{1\to2}^{\mathrm{mig}} - E_{2\to1}^{\mathrm{mig}} + E_{2\to3}^{\mathrm{mig}},$$

where the migration energies $E_{n\to n+1}^{\mathrm{mig}}$ and $E_{n\to n-1}^{\mathrm{mig}}$ have to be replaced by their values given in Tab. S5 for a $V_\alpha$ vacancy and in Tab. S7 for a $V_\beta$ vacancy.

So as to obtain the activation energy of the diffusion, we need to calculate the energy of formation of a vacancy in the $Al_3X$ $L1_2$ compound. This is given by

$$E_{V_\alpha}^{\mathrm{for}} = 8\varepsilon_{\mathrm{AlV}}^{(1)} + 4\varepsilon_{\mathrm{XV}}^{(1)} - 2\varepsilon_{\mathrm{AlAl}}^{(1)} - 4\varepsilon_{\mathrm{AlX}}^{(1)} + 6\varepsilon_{\mathrm{AlV}}^{(2)} - 3\varepsilon_{\mathrm{AlAl}}^{(2)},$$

for a $V_\alpha$ vacancy and by

$$E_{V_\beta}^{\mathrm{for}} = 12\varepsilon_{\mathrm{AlV}}^{(1)} - 3\varepsilon_{\mathrm{AlAl}}^{(1)} - 6\varepsilon_{\mathrm{AlX}}^{(1)} + 3\varepsilon_{\mathrm{XX}}^{(1)} + 6\varepsilon_{\mathrm{XV}}^{(2)} - 3\varepsilon_{\mathrm{XX}}^{(2)},$$

for a $V_\beta$ vacancy.

Another contribution arises from the interaction energy between the Y impurity and the vacancy lying in the configuration 0 of the cycle. For a $V_\alpha$ vacancy, this interaction energy is given by

$$E_{\mathrm{YV}_\alpha}^{\mathrm{inter}} = \varepsilon_{\mathrm{AlX}}^{(1)} + \varepsilon_{\mathrm{YV}}^{(1)} - \varepsilon_{\mathrm{AlY}}^{(1)} - \varepsilon_{\mathrm{XV}}^{(1)},$$

and for a $V_\beta$ vacancy by

$$E_{\mathrm{YV}_\beta}^{\mathrm{inter}} = \varepsilon_{\mathrm{XX}}^{(2)} + \varepsilon_{\mathrm{YV}}^{(2)} - \varepsilon_{\mathrm{XY}}^{(2)} - \varepsilon_{\mathrm{XV}}^{(2)}.$$

The activation energy associated to diffusion can now be calculated from the sum

$$E_{\mathrm{Y}}^{\mathrm{act}} = E_{\mathrm{Y}}^{\mathrm{mig}} + E_{\mathrm{V}}^{\mathrm{for}} + E_{\mathrm{YV}}^{\mathrm{inter}}.$$

For the diffusion of Sc inside $Al_3Zr$, the obtained activation energies are 6.34 eV in the case of a $V_\alpha$ vacancy and 5.62 eV in the case of a $V_\beta$ vacancy. These energies are considerably higher than the one corresponding to the diffusion of Sc in aluminium ($E_{\mathrm{Sc}}^{\mathrm{act}} = 1.79$ eV), thus showing that Sc cannot diffuse in $Al_3Zr$ precipitates. In the same way, high activation energies forbid any diffusion of Zr inside $Al_3Sc$ precipitates: we obtain $E_{\mathrm{Zr}}^{\mathrm{act}} = 5.51$ eV for a $V_\alpha$ vacancy and 5.26 eV for a $V_\beta$ vacancy whereas the activation energy of Zr diffusion in aluminium is only 2.51 eV.

## Sc – Vacancy Interaction

So as to compute the Sc – vacancy interaction $\varepsilon_{\mathrm{ScV}}^{(1)}$ we used the experimental binding energy $E_{\mathrm{ScV}}^{\mathrm{bin}}$ in aluminium deduced from electrical resistivity measurements[16]. These experiments show that there is a strong attraction between scandium atoms and vacancies and lead to $E_{\mathrm{ScV}}^{\mathrm{bin}} = 0.35$ eV. On the other hand, ab-initio calculations found repulsion when Sc atoms and vacancies are first nearest neighbours. Using the Kohn-Korringa-Rostoker Green's function method, Hoshino et al.[17] obtained $E_{\mathrm{ScV}}^{\mathrm{bin}} \approx -0.10$ eV and using LMTO calculations with the atomic sphere approximation, Sterne et al.[18] got $E_{\mathrm{ScV}}^{\mathrm{bin}} \approx -0.11$ eV. The clear contradiction between experiments and ab-initio calculations may suggest at a first glance that approximations used in these calculations were too crude: both calculations assumed spherical potentials and neglected relaxations. Nevertheless, recent ab-initio calculations[19] performed without such approximations confirmed the repulsion between Sc atoms and vacancies in aluminium: Sandberg and Holmestad obtained $E_{\mathrm{ScV}}^{\mathrm{bin}} \approx -0.24$ eV within local density approximation and $E_{\mathrm{ScV}}^{\mathrm{bin}} \approx -0.28$ eV within generalized gradient approximation. Some of the discrepancy may arise from the fact that ab-initio calculations were performed at 0 K whereas the experimental binding energy was measured between ~ 500 and 700 K, but one does not expect such a big difference from a temperature dependence. This suggests that the binding energy deduced from experiments might be wrong and reveals the necessity of performing new experiments so as to confirm or invalidate the previous ones.

One can wonder what would have happened in our atomic simulations if we had used the ab-initio binding energy $E_{\mathrm{ScV}}^{\mathrm{bin}}$ instead of the experimental one. In this purpose, we build another atomic diffusion model using the value $E_{\mathrm{ScV}}^{\mathrm{bin}} \approx -0.28$ calculated by Sandberg and Holmestad. The only parameter that needs to be changed in the atomic model is the interaction between a Sc atom and a vacancy when they are first nearest neighbours which, now, takes the value $\varepsilon_{\mathrm{ScV}}^{(1)} = -0.141\,\mathrm{eV}$. All other parameters remain unchanged. In particular, we do not modify the contribution $e_{\mathrm{Sc}}^{\mathrm{sp}}$ of Sc atoms to the saddle point energy which is fitted to the activation energy of Sc impurity diffusion in aluminium thanks to the equation

$$E_{\mathrm{Sc}}^{\mathrm{act}}(\mathrm{Al}) = e_{\mathrm{Sc}}^{\mathrm{sp}} - 5\varepsilon_{\mathrm{AlAl}}^{(1)} - 12\varepsilon_{\mathrm{AlSc}}^{(1)} - 6\varepsilon_{\mathrm{AlSc}}^{(2)} .$$

This means that the activation energy for diffusion is unchanged but split differently between the interaction energy of Sc atoms with vacancies and the Sc migration energy. Using the six-jump vacancy cycle model described above, we can show that this new atomic model leads to the same conclusion as the one we use in our study: precipitates cannot rearrange themselves to reach their equilibrium structures. Indeed, the values obtained for the activation energies of the diffusion of Sc in $Al_3Zr$ (6.34 eV for a $V_\alpha$ vacancy and 5.62 eV for $V_\beta$) and of Zr in $Al_3Sc$ (5.51 eV for a $V_\alpha$ vacancy and 5.26 eV for $V_\beta$) are still considerably larger than activation energies corresponding to solute diffusion in aluminium. We confirm this prediction by running kinetic Monte Carlo simulations with this new set of parameters. They lead to precipitates showing the same heterogeneous structure as the one displayed in Fig. 1 in the manuscript or in Fig. S4. It thus demonstrates that this structure does not depend on the interaction between Sc atoms and vacancies but solely on the difference between diffusion coefficients of both solute atoms and on the absence of solute diffusion inside precipitates. Our simulations cannot then be used to discriminate between both values proposed for the binding energy between Sc atoms and vacancies, the ab-initio and the experimental one.

## References for supplementary information

1. Clouet, E., Nastar, M. & Sigli, C. Nucleation of $Al_3Zr$ and $Al_3Sc$ in aluminum alloys: from kinetic Monte Carlo simulations to classical theory. *Phys. Rev. B* **69**, 064109 (2004).
2. Ducastelle, F. *Order and Phase Stability in Alloys* (North-Holland, Amsterdam, 1991).
3. Clouet, E., Sanchez J. M. & Sigli, C. First-principles study of the solubility of Zr in Al. *Phys. Rev. B* **65**, 094105 (2002).
4. Clouet, E. *Séparation de phase dans les alliages Al-Zr-Sc: du saut des atomes à la croissance de précipités ordonnés* (PhD thesis, Ecole Centrale Paris, 2004) ; http://tel.ccsd.cnrs.fr/tel-00005967.
5. Clouet, E., Barbu, A., Laé, L. & Martin, G. Precipitation kinetics of $Al_3Zr$ and $Al_3Sc$ in aluminum alloys modeled with cluster dynamics. *Acta Mater.* **53**, 2313-2325 (2005).
6. Clouet E., Nastar. N., Barbu A., Sigli C. & Martin G. Precipitation in Al-Zr-Sc alloys: a comparison between kinetic Monte Carlo, cluster dynamics and classical nucleation theory. In : *Solid-Solid Phase Transformations in Inorganic Materials*. Howe, J. M., Laughlin, D. E., Lee, J. K., Dahmen, U. & Soffa, W. A. ed. **2**, 683-703 (TMS, 2005) ; http://xxx.lanl.gov/abs/cond-mat/0507259.
7. Methfessel, M. & van Schilfgaarde, M. Derivation of force theorem in density-functional theory: application to the full-potential LMTO method. *Phys. Rev. B* **48**, 4987-4940 (1993).
8. Perdew, J. P., Burke, K. & Ernzerhof, M. Generalized gradient approximation made simple. *Phys. Rev. Lett.* **77**, 3865-3868 (1996).
9. Xu, J. H. & Freeman, A. J. Phase stability and electronic structures of $ScAl_3$ and $ZrAl_3$ and of Sc-stabilized cubic $ZrAl_3$ precipitates. *Phys. Rev. B* **41**, 12553-12561 (1990).
10. Yelagin, V. I., Zakharov, V. V., Pavlenko, S. G. & Rostova, T. D. Influence of zirconium additions on ageing of Al-Sc alloys. *Phys. Met. Metall.* **60**, 88-92 (1985).
11. Fuller, C. B., Seidman, D. N. & Dunand, D. C. Mechanical properties of Al(Sc,Zr) alloys at ambient and elevated temperatures. *Acta Mater.* **51**, 4803-4814 (2003).
12. Riddle, Y. W. & Sanders, T. H. A study of coarsening, recrystallization and morphology of microstructure in Al-Sc-(Zr)-(Mg) alloys. *Metal. Mater. Trans. A* **35**, 341-350 (2004).
13. Young, W. M. & Elcock, E. W. Monte Carlo studies of vacancy migration in binary ordered alloys: I. *Proc. Phys. Soc.* **89**, 735-746 (1966).
14. Maeda, S., Tanaka, K. & Koiwa M. Diffusion via six-jump vacancy cycles in the $L1_2$ lattice. *Defect Diff. Forum* **95-98**, 855-858 (1993).
15. Athènes, M. & Bellon, P. Antisite-assisted diffusion in the $L1_2$ ordered structure studied by Monte Carlo simulations. *Phil. Mag. A* **79**, 2243-2257 (1999).
16. Miura, Y., Joh, C.-H. & Katsube, T. Determination of vacancy-Sc interaction energy by electrical resistivity measurements. *Mater. Sci. Forum* **331-337**, 1031-1036 (2000).
17. Hoshino, T., Zeller, R. & Dederichs P. H. Local-density-functional calculations for defect interactions in Al. *Phys. Rev. B* **53**, 8971-8974 (1996).
18. Sterne, P. A., van Ek, J. & Howell R. H. Electronic structure calculations of vacancies and their influence on materials properties. *Comput. Mater. Sci.* **10**, 306-313 (1998).
19. Sandberg, N. & Holmestad, R. First-principles calculations of impurity diffusion activation energies in Al. *Phys. Rev. B* **73**, 014108 (2006).

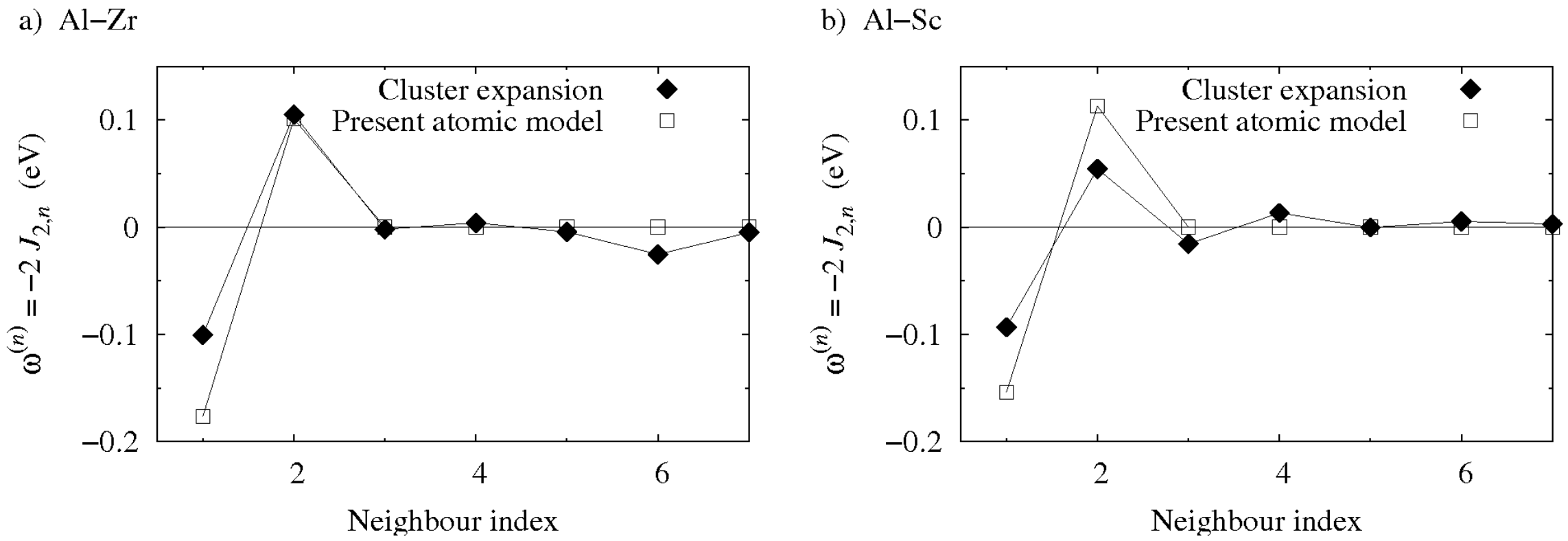

Figure S1 **Order energies.** The dependence with the neighbour index *n* of the order energies $\omega^{(n)} = \varepsilon_{ij}^{(n)} - (1/2)\varepsilon_{ii}^{(n)} - (1/2)\varepsilon_{jj}^{(n)}$ in the binary alloys Al-Zr (a) and Al-Sc (b) used in the present atomic model is compared with the one obtained from a cluster expansion of ab-initio calculations. For Al-Zr alloy, the cluster expansion is given in Ref. 3 whereas for Al-Sc alloy, it is given in the appendix A of Ref. 4.

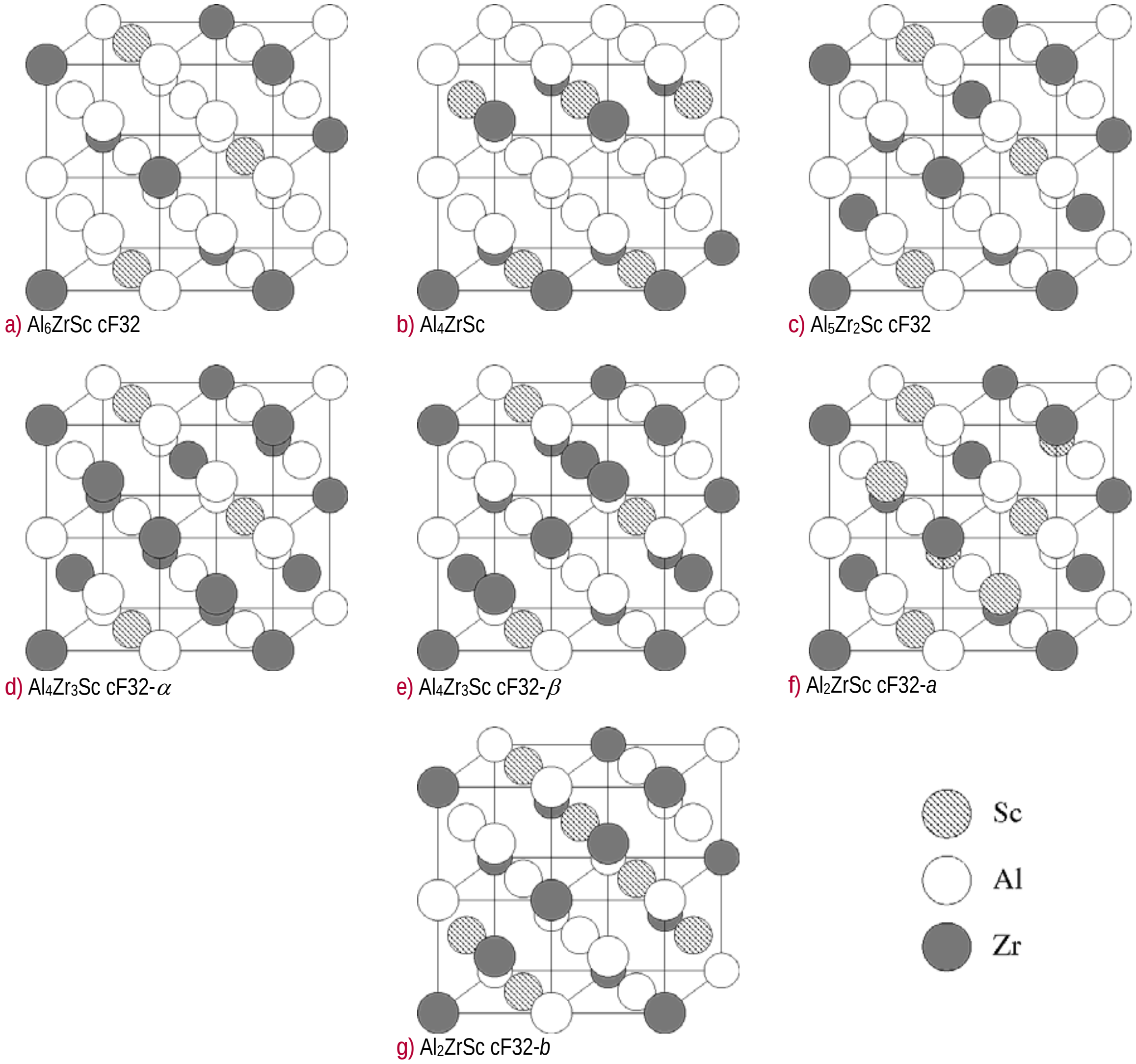

**Figure S2 $Al_xZr_ySc_z$ compounds.** Ternary ordered compounds used to fit the first nearest neighbour order interaction $\omega^{(1)}_{\mathrm{ZrSc}} = \varepsilon^{(1)}_{\mathrm{ZrSc}} - (1/2)\varepsilon^{(1)}_{\mathrm{ZrZr}} - (1/2)\varepsilon^{(1)}_{\mathrm{ScSc}}$ .

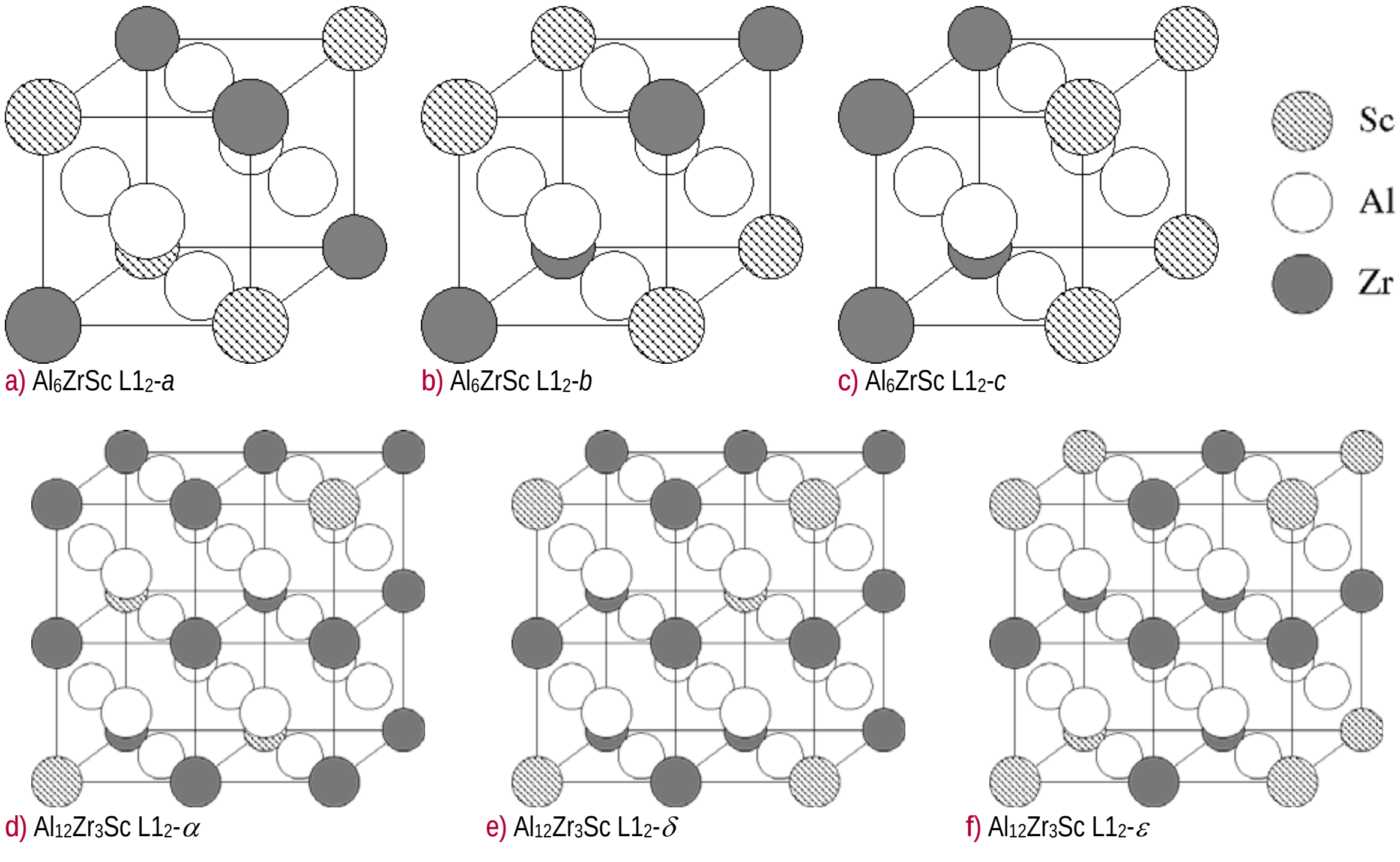

**Figure S3 $Al_3Zr_xSc_{1-x}$ compounds.** Ternary ordered compounds used to fit the second nearest neighbour order interaction $\omega^{(2)}_{ZrSc} = \varepsilon^{(2)}_{ZrSc} - (1/2)\varepsilon^{(2)}_{ZrZr} - (1/2)\varepsilon^{(2)}_{ScSc}$ .

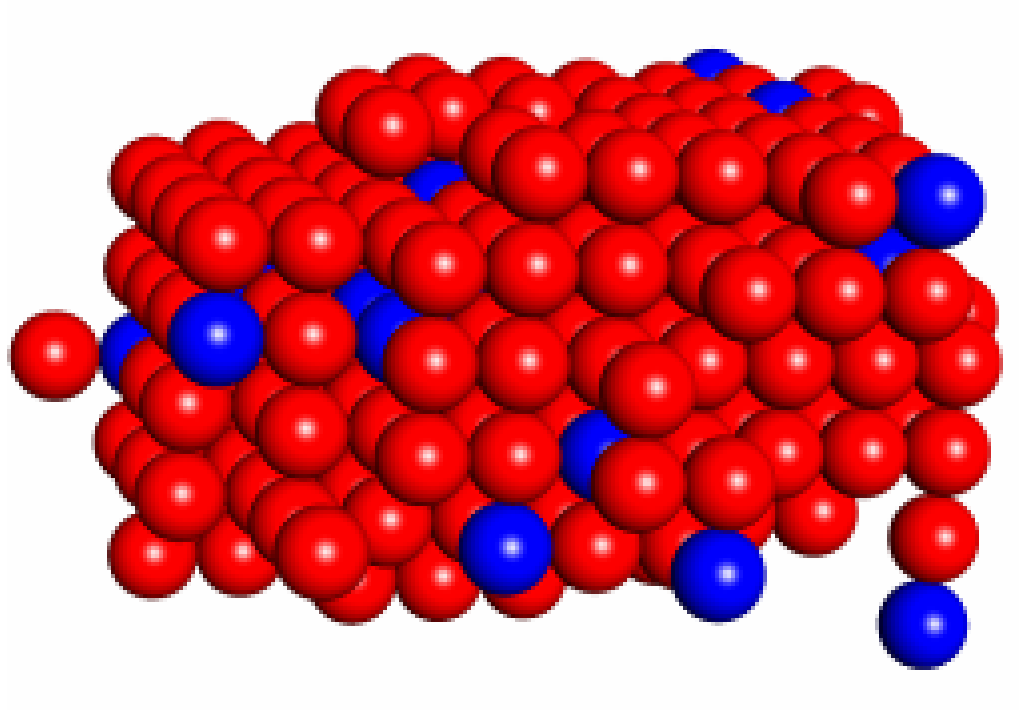

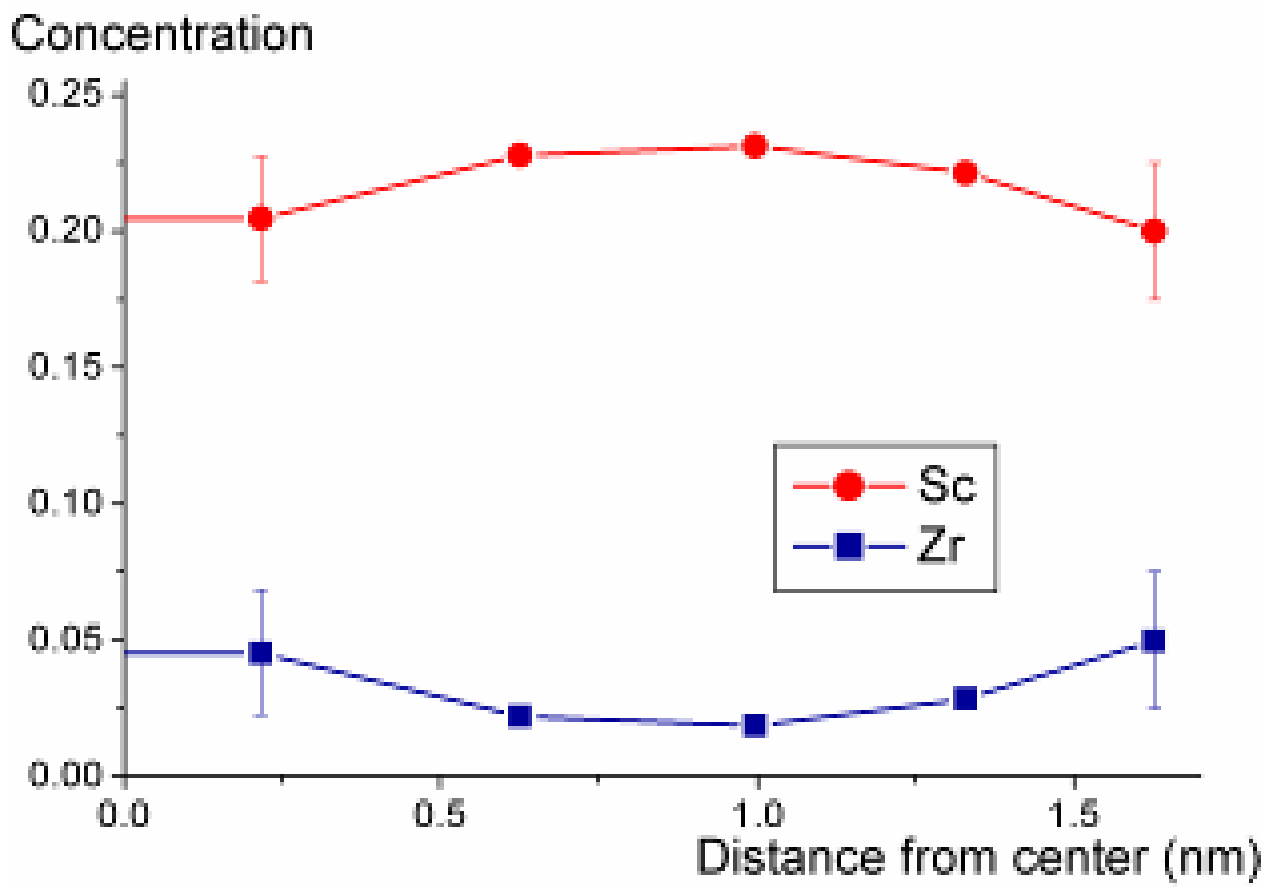

a) 5920 s at 450°C.

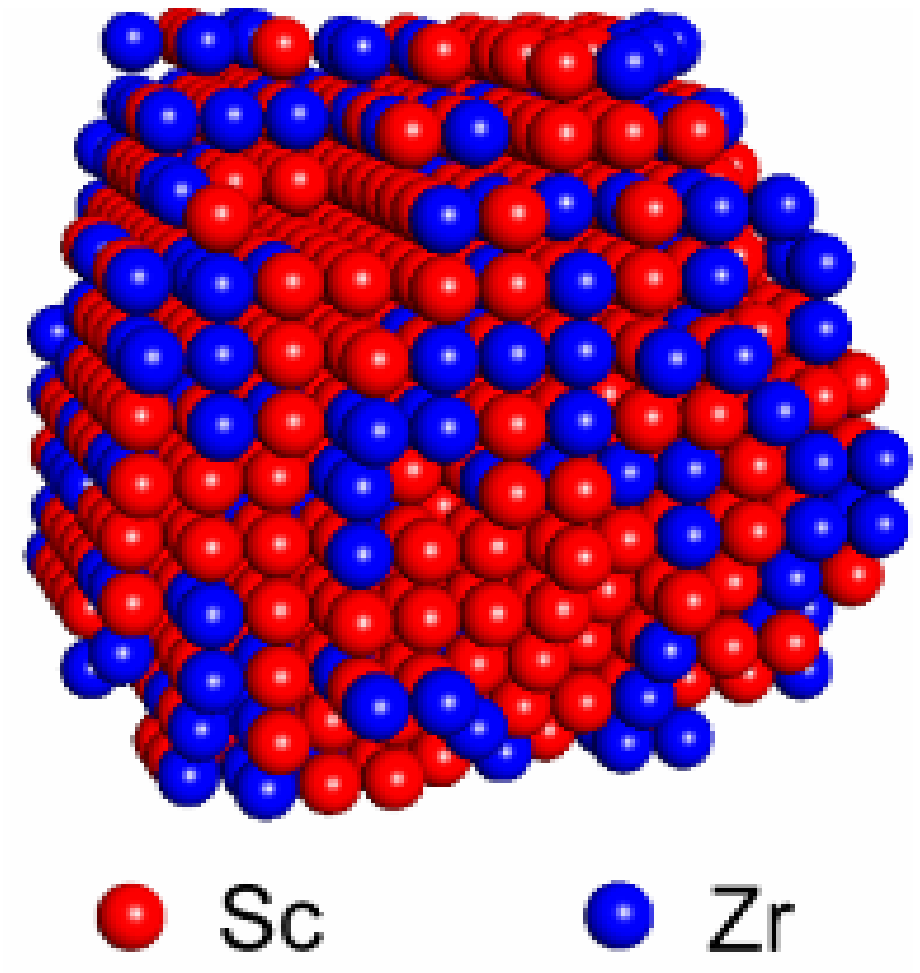

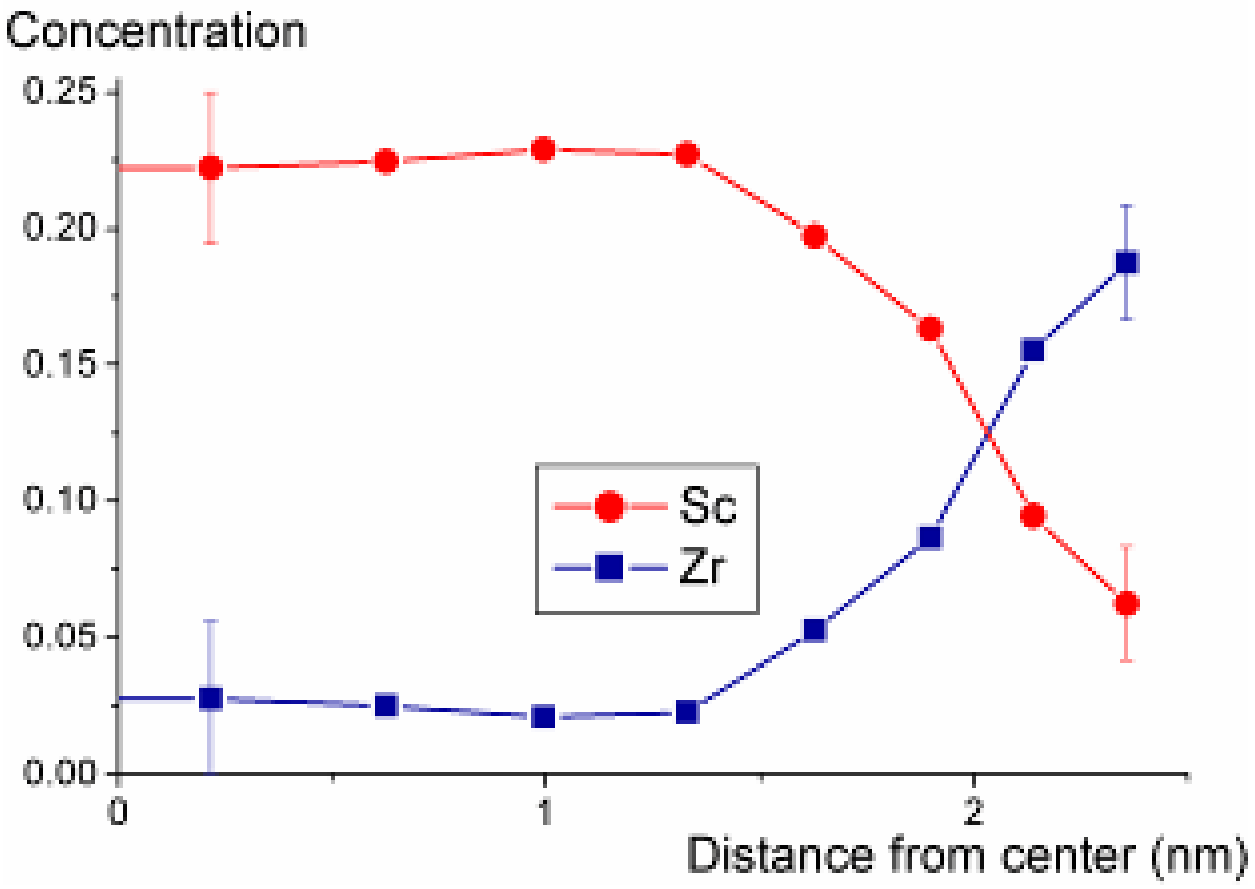

b) 0.59 s at 550°C.

Figure S4 $Al_3Zr_xSc_{1-x}$ **precipitates obtained by atomic simulations.** A precipitate has been isolated in the computational box after the simulated annealing at 450°C during 5920 s (a) and at 550°C during 0.59 s (b) of an aluminium solid solution containing 0.5 at.% Zr and 0.5 at.% Sc. The radial concentration profile shows the Zr enrichment of the periphery compared to the core.

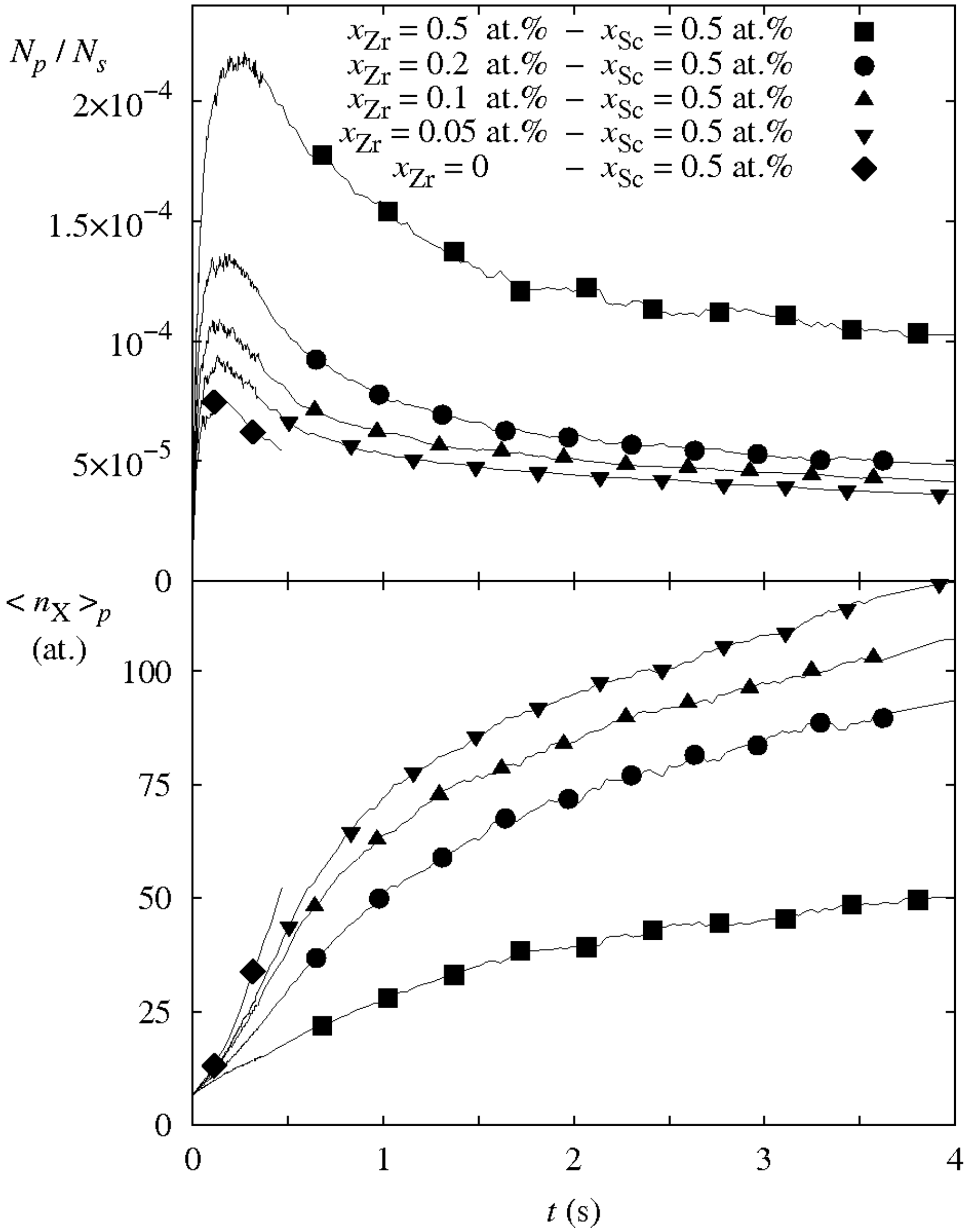

Figure S5 **Kinetics of precipitation.** Variation with the Zr concentration of the number $N_p$ or precipitates (normalized by the number $N_s$ of lattice sites) and of their mean size $< n_X >_p$ in KMC simulations of the annealing at 450°C of solid solutions containing 0.5 at.% Sc. The critical size used to discriminate precipitates from sub-critical clusters is $n_X^* = 6$ atoms.

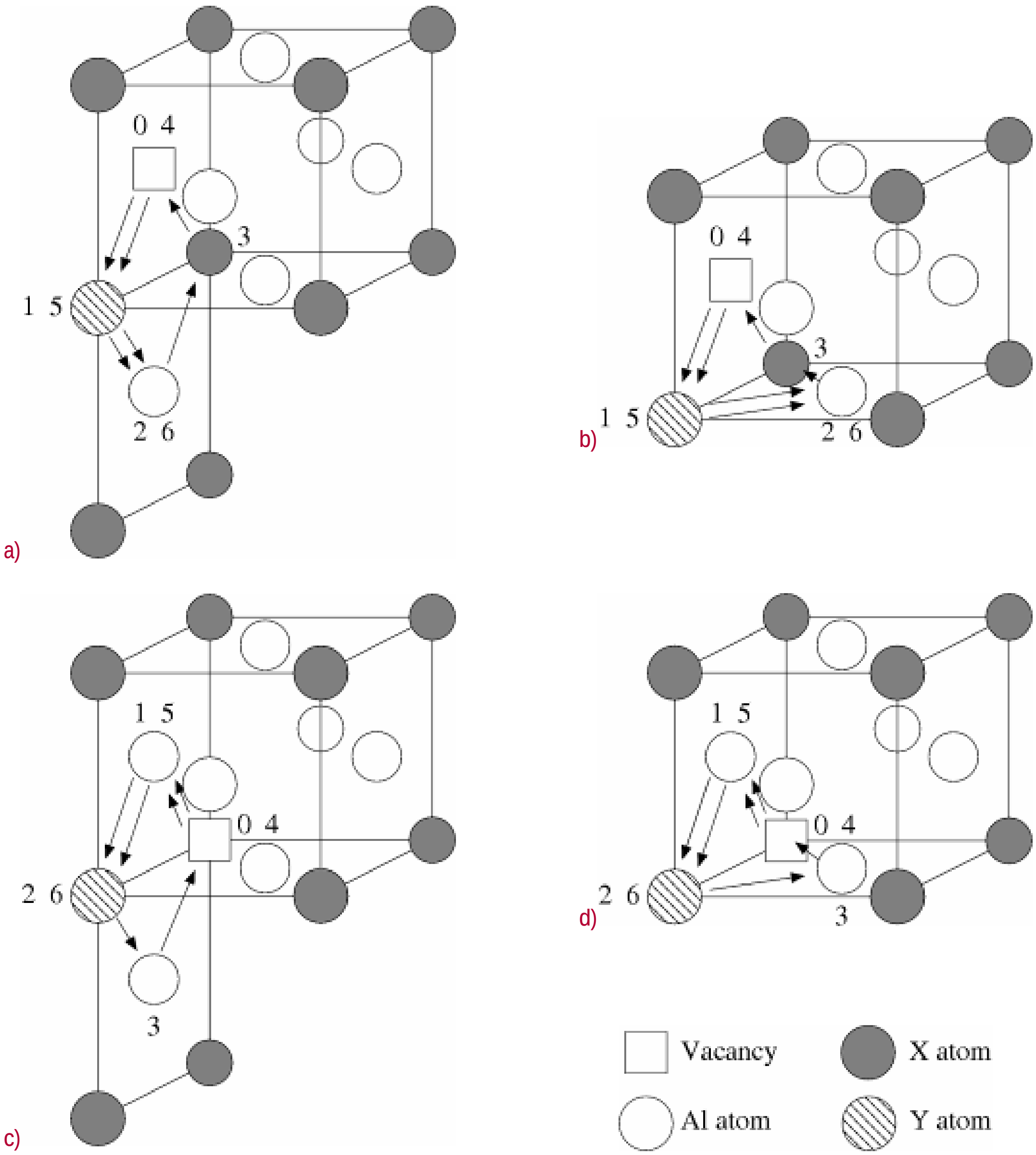

Figure S6 **Six-jump cycles in the $L1_2$ structure.** Diffusion of a Y impurity in the $L1_2$ structure of the $Al_3X$ compound starting from a vacancy on the majority (a,b) or minority (c,d) sub-lattice. (a) and (c) corresponds to a straight cycle and (b) and (d) to a bent cycle. The numbers indicate the vacancy position after the corresponding vacancy jump.

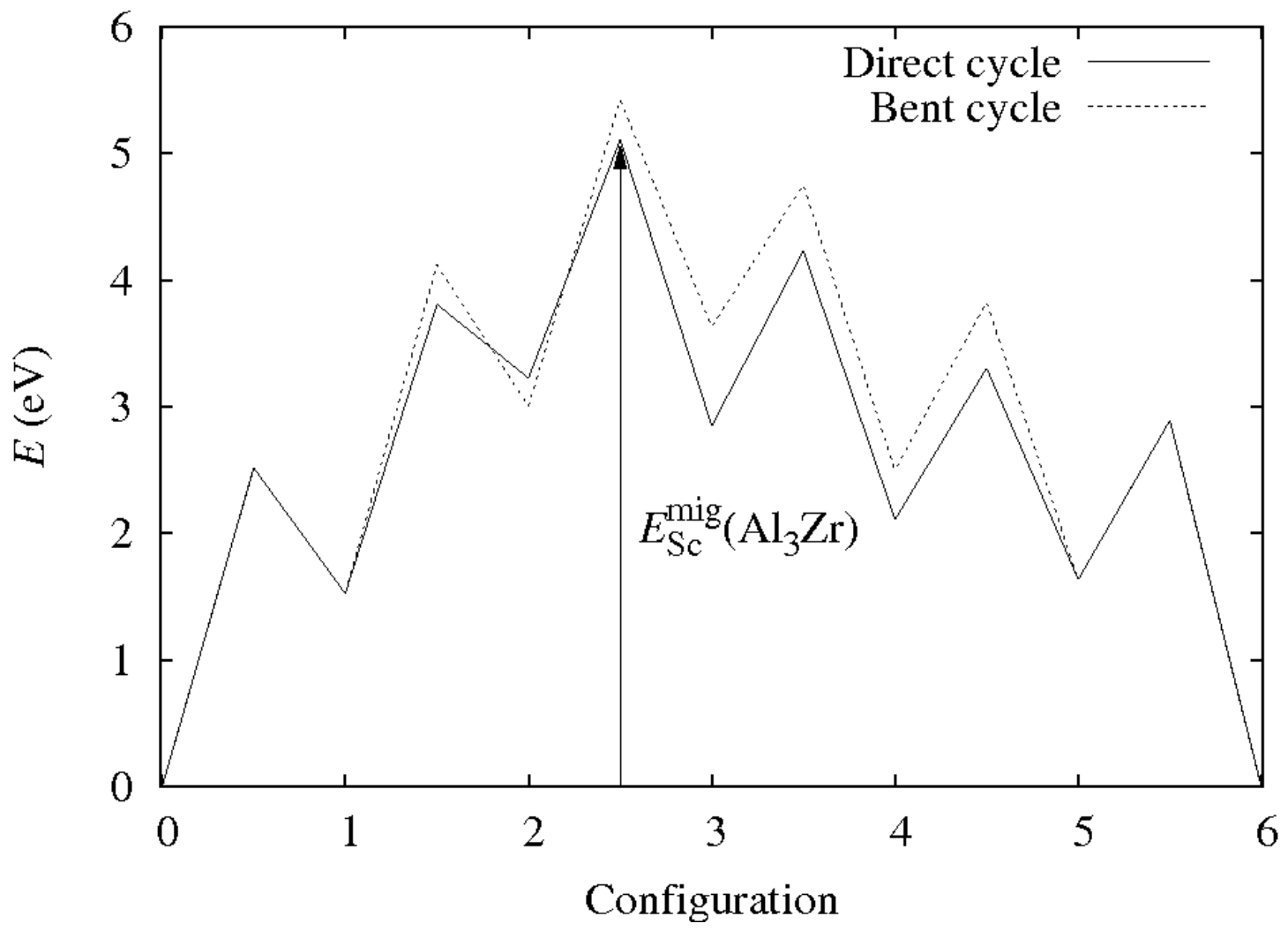

a)

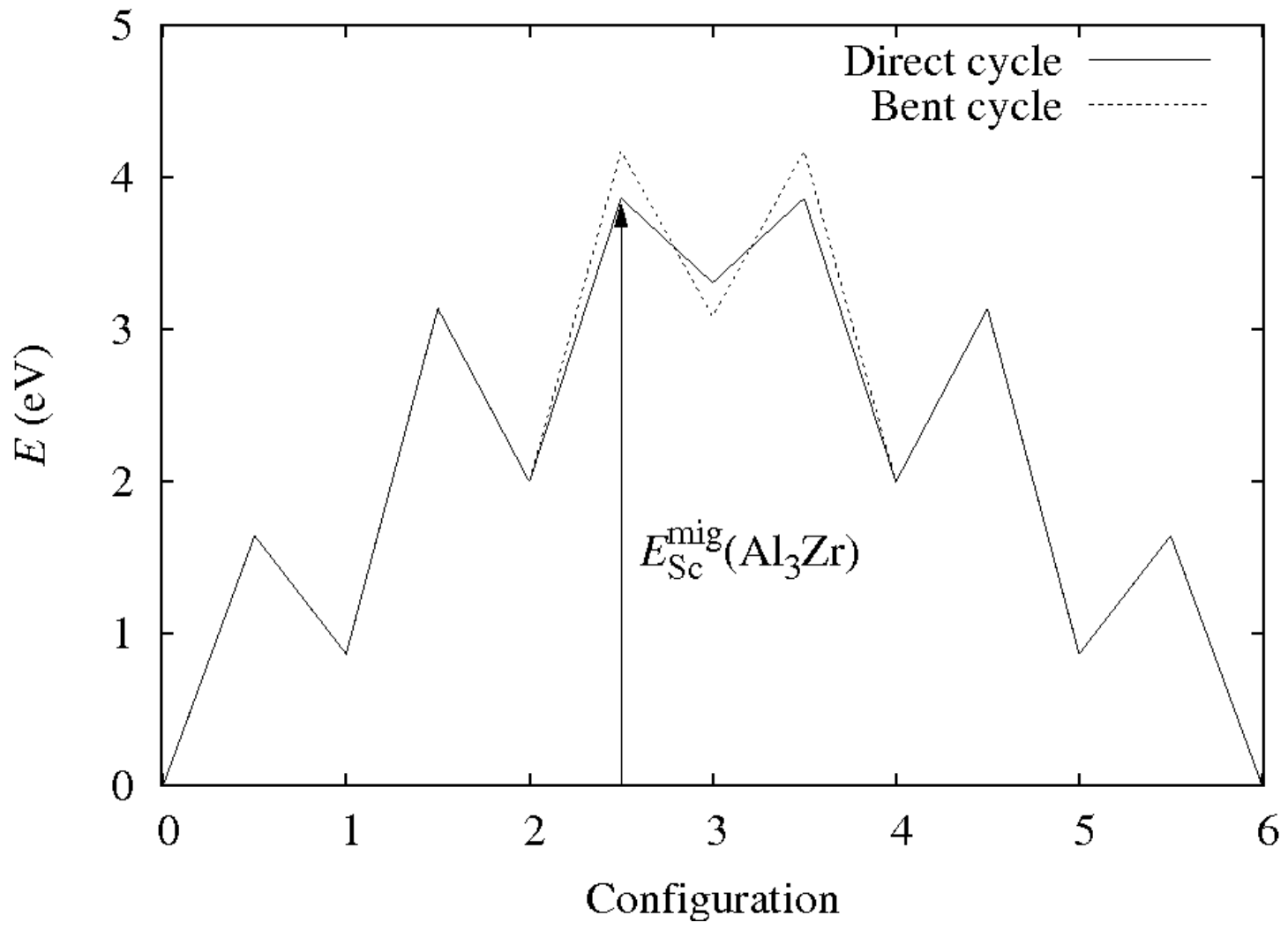

b)

Figure S7 **Diffusion of Sc in the $L1_2$ structure of $Al_3Zr$ via six-jump cycles**. The configuration and activation energies corresponding to the local minima and maxima of the energy landscape are deduced from the atomic diffusion model. The plots (a) and (b) respectively correspond to cycles starting with a vacancy on the majority or minority sub-lattice.

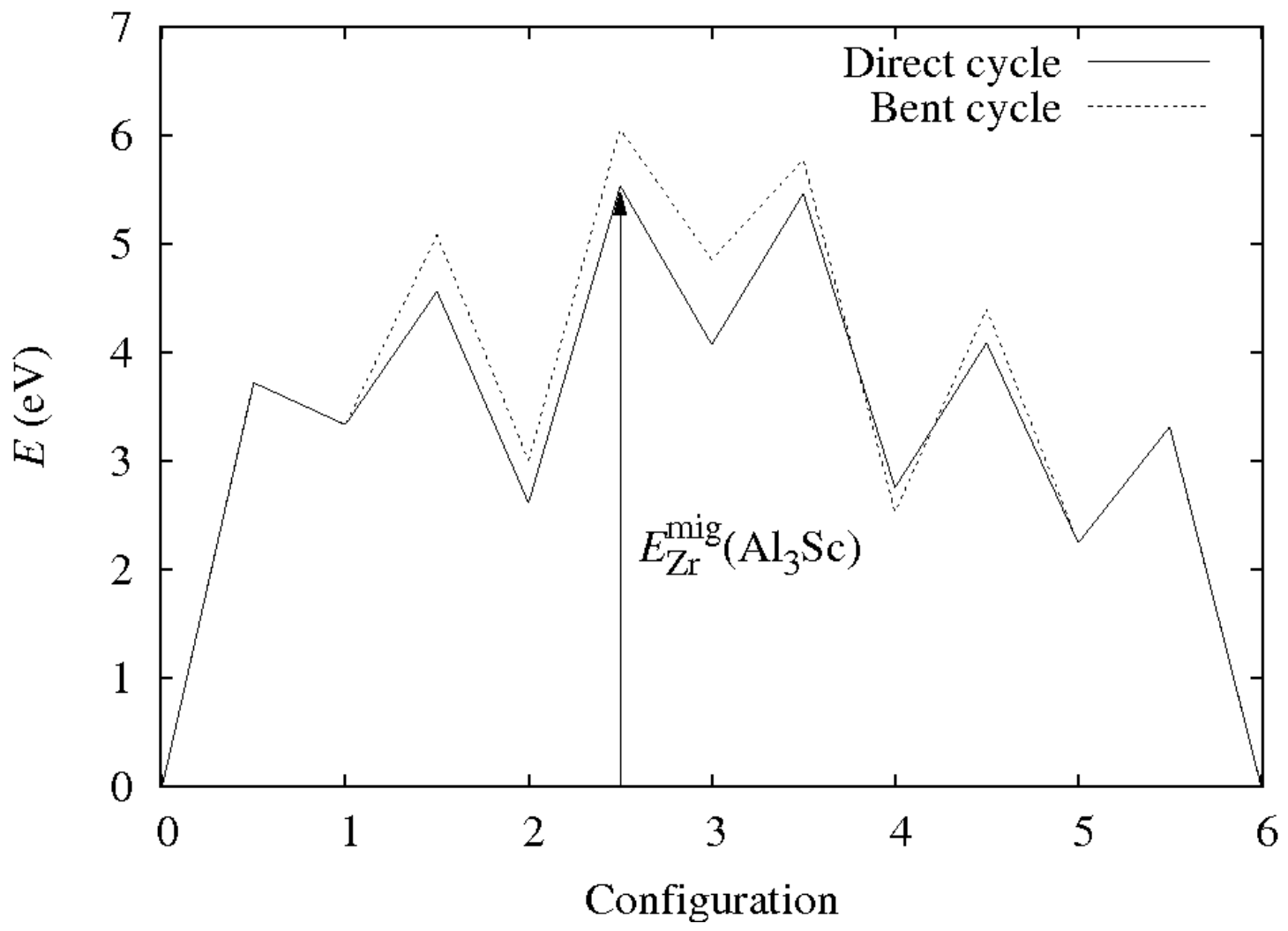

a)

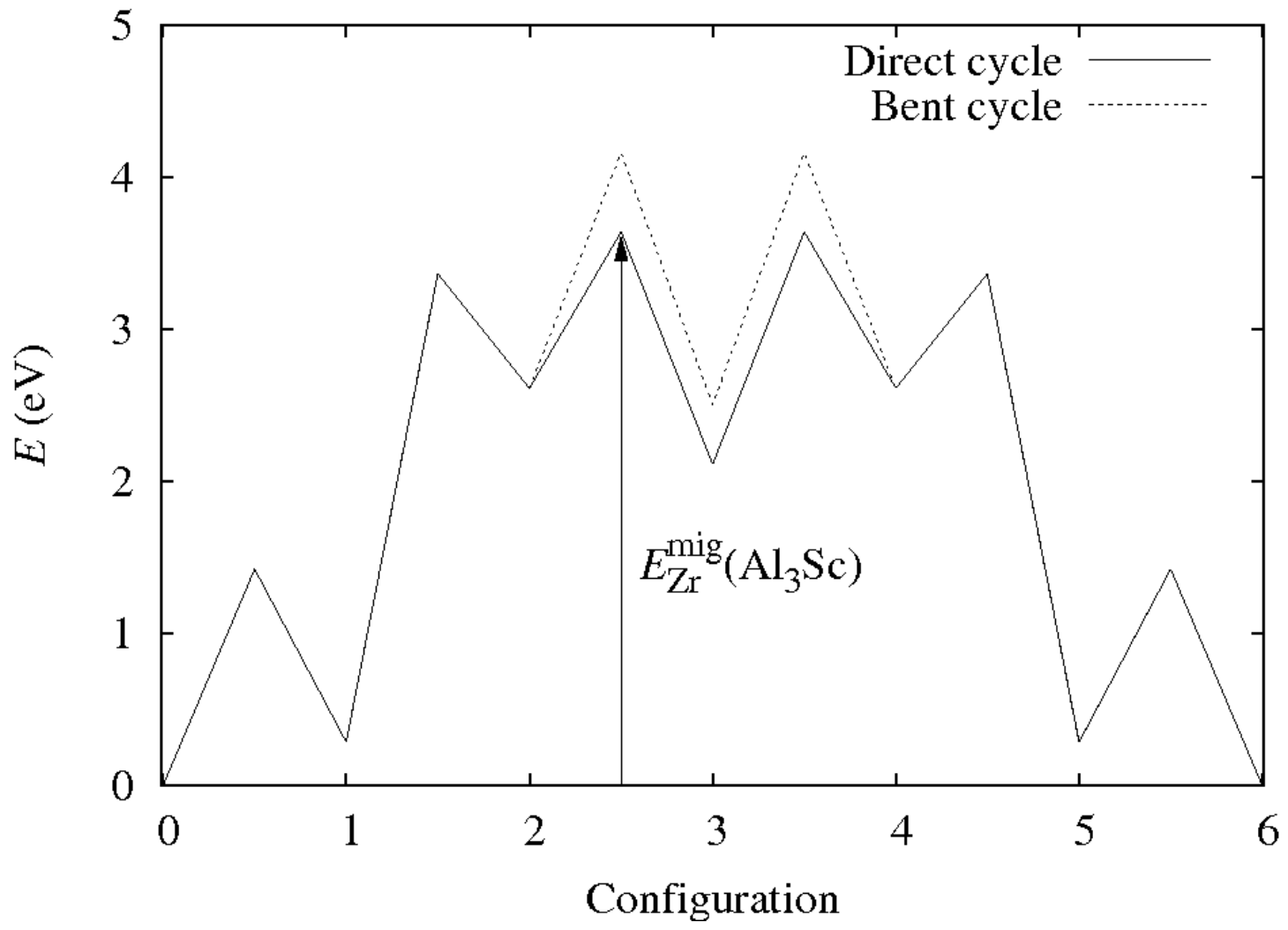

b)

Figure S8 **Diffusion of Zr in the $L1_2$ structure of $Al_3Sc$ via six-jump cycles.** The configuration and activation energies corresponding to the local minima and maxima of the energy landscape are deduced from the atomic diffusion model. The plots (a) and (b) respectively correspond to cycles starting with a vacancy on the majority or minority sub-lattice.

| | | $\Delta E^{\text{form}}$ (meV/at.) | |
|---|---|---|---|
| | | Ab-initio | Ising |
| $Al_6Zr$ | cF32 | -238 | -193 |
| $Al_4ZrSc$ | | -184 | -211 |
| $Al_5Zr_2Sc$ | cF32 | -141 | -152 |
| $Al_5ZrSc_2$ | cF32 | -136 | -121 |
| $Al_4Zr_3Sc$ | cF32-$\alpha$ | 5 | -23 |
| $Al_4Zr_3Sc$ | cF32-$\beta$ | 15 | -23 |
| $Al_2ZrSc$ | cF32-$a$ | 12 | 62 |
| $Al_4Zr_2Sc_2$ | cF32-$b$ | 39 | 62 |
| $Al_4ZrSc_3$ | cF32-$\alpha$ | 23 | 29 |
| $Al_4ZrSc_3$ | cF32-$\beta$ | 56 | 29 |

Table S1 Formation energies of $Al_xZr_ySc_z$ ordered compounds used to fit the first nearest neighbour interaction $\omega^{(1)}_{\text{ZrSc}} = \varepsilon^{(1)}_{\text{ZrSc}} - (1/2)\varepsilon^{(1)}_{\text{ZrZr}} - (1/2)\varepsilon^{(1)}_{\text{ScSc}}$. The reference states used to define the formation energies are the fcc structures of Al, Zr and Sc at their equilibrium volume: $\Delta E^{\text{form}}(Al_xZr_ySc_z) = E(Al_xZr_ySc_z) - x\,E(Al) - y\,E(Zr) - z\,E(Sc)$. All ab-initio energies of the compounds are calculated at the equilibrium volume of Al ($V_0$=16.53 Å$^3$). The standard deviation between ab-initio calculations and the Ising model is $\delta E$ = 30 meV/at.

| | | $V_0$ (Å$^3$/at.) | $\Delta E$ (meV/at.) | |
|---|---|---|---|---|
| | | Ab-initio | Ab-initio | Ising |
| $Al_3Zr$ | $L1_2$ | 16.89 | 0. | 0. |
| $Al_{12}Zr_3Sc$ | $L1_2$-$\alpha$ | 17.01 | -1.90 | -4.15 |
| $Al_{12}Zr_3Sc$ | $L1_2$-$\delta$ | 17.01 | -4.35 | -4.15 |
| $Al_{12}Zr_3Sc$ | $L1_2$-$\varepsilon$ | 17.01 | -2.45 | -2.76 |
| $Al_6ZrSc$ | $L1_2$-$a$ | 17.09 | -5.44 | -8.30 |
| $Al_6ZrSc$ | $L1_2$-$b$ | 17.11 | -8.16 | -5.58 |
| $Al_6ZrSc$ | $L1_2$-$c$ | 17.10 | -4.22 | -2.72 |
| $Al_{12}ZrSc_3$ | $L1_2$-$\alpha$ | 17.19 | -3.26 | -4.15 |
| $Al_{12}ZrSc_3$ | $L1_2$-$\delta$ | 17.18 | -6.67 | -4.15 |
| $Al_{12}ZrSc_3$ | $L1_2$-$\varepsilon$ | 17.19 | -5.17 | -2.76 |
| $Al_3Sc$ | $L1_2$ | 17.28 | 0. | 0. |

Table S2 Energies of $Al_3Zr_xSc_{1-x}$ ordered compounds used to fit the second nearest neighbour interaction $\omega^{(2)}_{\text{ZrSc}} = \varepsilon^{(2)}_{\text{ZrSc}} - (1/2)\varepsilon^{(2)}_{\text{ZrZr}} - (1/2)\varepsilon^{(2)}_{\text{ScSc}}$. The reference states used to define the energies are the $Al_3Zr$ and $Al_3Sc$ compounds in their $L1_2$ structure: $\Delta E(Al_3Zr_xSc_{1-x}) = E(Al_3Zr_xSc_{1-x}) - x\,E(Al_3Zr) - (1-x)\,E(Al_3Sc)$. All ab-initio energies are calculated at the equilibrium volume $V_0$ of the given structure. The standard deviation between ab-initio calculations and the Ising model is $\delta E$ = 2 meV/at.

| | |
|---|---|
| $\varepsilon_{\mathrm{AlAl}}^{(1)} = -560$ | $\varepsilon_{\mathrm{AlAl}}^{(2)} = 0$ |
| $\varepsilon_{\mathrm{AlZr}}^{(1)} = -979 + 0.244T$ | $\varepsilon_{\mathrm{AlZr}}^{(2)} = 101 - 0.0233T$ |
| $\varepsilon_{\mathrm{AlSc}}^{(1)} = -759 + 0.21T$ | $\varepsilon_{\mathrm{AlSc}}^{(2)} = 113 - 0.0334T$ |
| $\varepsilon_{\mathrm{ZrZr}}^{(1)} = -1045$ | $\varepsilon_{\mathrm{ZrZr}}^{(2)} = 0$ |
| $\varepsilon_{\mathrm{ScSc}}^{(1)} = -650$ | $\varepsilon_{\mathrm{ScSc}}^{(2)} = 0$ |
| $\varepsilon_{\mathrm{ZrSc}}^{(1)} = -611$ | $\varepsilon_{\mathrm{ZrSc}}^{(2)} = -0.00277$ |
| $\varepsilon_{\mathrm{AlV}}^{(1)} = -222$ | $\varepsilon_{\mathrm{AlV}}^{(2)} = 0$ |
| $\varepsilon_{\mathrm{ZrV}}^{(1)} = -350$ | $\varepsilon_{\mathrm{ZrV}}^{(2)} = 0$ |
| $\varepsilon_{\mathrm{ScV}}^{(1)} = -757$ | $\varepsilon_{\mathrm{ScV}}^{(2)} = 0$ |
| $\varepsilon_{\mathrm{VV}}^{(1)} = 0$ | $\varepsilon_{\mathrm{VV}}^{(2)} = 0$ |

Table S3 First and second nearest neighbour pair interactions. Energies are in meV and temperature $T$ in K.

| | |
|---|---|
| $e_{\mathrm{Al}}^{sp} = -8.219$ eV | $\nu_{\mathrm{Al}} = 1.36 \times 10^{14}$ Hz |
| $e_{\mathrm{Zr}}^{sp} = -11.464$ eV | $\nu_{\mathrm{Zr}} = 9 \times 10^{16}$ Hz |
| $e_{\mathrm{Sc}}^{sp} = -9.434$ eV | $\nu_{\mathrm{Sc}} = 4 \times 10^{15}$ Hz |

Table S4 Kinetic parameters: contribution of the jumping atom to the saddle point energy $e_{\alpha}^{sp}$ and attempt frequency $\nu_{\alpha}$ for $\alpha \equiv$ Al, Zr, and Sc atoms.

| Forward jumps |
|---|
| $E_{0\to1}^{\text{mig}} = e_{\text{Y}}^{\text{sp}} - 11\varepsilon_{\text{AlY}}^{(1)} - 8\varepsilon_{\text{AlV}}^{(1)} - 3\varepsilon_{\text{XV}}^{(1)} - \varepsilon_{\text{YV}}^{(1)} - 6\varepsilon_{\text{XY}}^{(2)} - 6\varepsilon_{\text{AlV}}^{(2)}$ |
| $E_{1\to2}^{\text{mig}} = e_{\text{Al}}^{\text{sp}} - 8\varepsilon_{\text{AlAl}}^{(1)} - 3\varepsilon_{\text{AlX}}^{(1)} - 11\varepsilon_{\text{AlV}}^{(1)} - \varepsilon_{\text{YV}}^{(1)} - 5\varepsilon_{\text{AlAl}}^{(2)} - \varepsilon_{\text{AlY}}^{(2)} - 6\varepsilon_{\text{XV}}^{(2)}$ |
| $E_{2\to3}^{\text{mig}} = e_{\text{X}}^{\text{sp}} - 10\varepsilon_{\text{AlX}}^{(1)} - \varepsilon_{\text{XY}}^{(1)} - 9\varepsilon_{\text{AlV}}^{(1)} - 3\varepsilon_{\text{XV}}^{(1)} - \varepsilon_{\text{AlX}}^{(2)} - 5\varepsilon_{\text{XX}}^{(2)} - 5\varepsilon_{\text{AlV}}^{(2)} - \varepsilon_{\text{YV}}^{(2)}$ |
| $E_{3\to4}^{\text{mig}} = e_{\text{Y}}^{\text{sp}} - 9\varepsilon_{\text{AlY}}^{(1)} - 2\varepsilon_{\text{XY}}^{(1)} - 10\varepsilon_{\text{AlV}}^{(1)} - \varepsilon_{\text{XV}}^{(1)} - \varepsilon_{\text{YV}}^{(1)} - 5\varepsilon_{\text{AlY}}^{(2)} - \varepsilon_{\text{XY}}^{(2)} - \varepsilon_{\text{AlV}}^{(2)} - 5\varepsilon_{\text{XV}}^{(2)}$ |
| $E_{4\to5}^{\text{mig}} = e_{\text{Al}}^{\text{sp}} - 10\varepsilon_{\text{AlAl}}^{(1)} - \varepsilon_{\text{AlX}}^{(1)} - 9\varepsilon_{\text{AlV}}^{(1)} - 2\varepsilon_{\text{XV}}^{(1)} - \varepsilon_{\text{YV}}^{(1)} - 5\varepsilon_{\text{AlX}}^{(2)} - \varepsilon_{\text{AlY}}^{(2)} - 5\varepsilon_{\text{AlV}}^{(2)} - \varepsilon_{\text{XV}}^{(2)}$ |
| $E_{5\to6}^{\text{mig}} = e_{\text{X}}^{\text{sp}} - 8\varepsilon_{\text{AlX}}^{(1)} - 2\varepsilon_{\text{XX}}^{(1)} - \varepsilon_{\text{XY}}^{(1)} - 11\varepsilon_{\text{AlV}}^{(1)} - \varepsilon_{\text{XV}}^{(1)} - 6\varepsilon_{\text{AlX}}^{(2)} - 5\varepsilon_{\text{XV}}^{(2)} - \varepsilon_{\text{YV}}^{(2)}$ |
| **Backward jumps** |
| $E_{1\to0}^{\text{mig}} = e_{\text{Y}}^{\text{sp}} - 8\varepsilon_{\text{AlY}}^{(1)} - 3\varepsilon_{\text{XY}}^{(1)} - 11\varepsilon_{\text{AlV}}^{(1)} - \varepsilon_{\text{YV}}^{(1)} - 6\varepsilon_{\text{AlY}}^{(2)} - 6\varepsilon_{\text{XV}}^{(2)}$ |
| $E_{2\to1}^{\text{mig}} = e_{\text{Al}}^{\text{sp}} - 10\varepsilon_{\text{AlAl}}^{(1)} - \varepsilon_{\text{AlY}}^{(1)} - 9\varepsilon_{\text{AlV}}^{(1)} - 3\varepsilon_{\text{XV}}^{(1)} - 6\varepsilon_{\text{AlX}}^{(2)} - 5\varepsilon_{\text{AlV}}^{(2)} - \varepsilon_{\text{YV}}^{(2)}$ |
| $E_{3\to2}^{\text{mig}} = e_{\text{X}}^{\text{sp}} - 9\varepsilon_{\text{AlX}}^{(1)} - 2\varepsilon_{\text{XX}}^{(1)} - 10\varepsilon_{\text{AlV}}^{(1)} - \varepsilon_{\text{XV}}^{(1)} - \varepsilon_{\text{YV}}^{(1)} - 5\varepsilon_{\text{AlX}}^{(2)} - \varepsilon_{\text{XY}}^{(2)} - \varepsilon_{\text{AlV}}^{(2)} - 5\varepsilon_{\text{XV}}^{(2)}$ |
| $E_{4\to3}^{\text{mig}} = e_{\text{Y}}^{\text{sp}} - 10\varepsilon_{\text{AlY}}^{(1)} - \varepsilon_{\text{XY}}^{(1)} - 9\varepsilon_{\text{AlV}}^{(1)} - 2\varepsilon_{\text{XV}}^{(1)} - \varepsilon_{\text{YV}}^{(1)} - \varepsilon_{\text{AlY}}^{(2)} - 5\varepsilon_{\text{XY}}^{(2)} - 5\varepsilon_{\text{AlV}}^{(2)} - \varepsilon_{\text{XV}}^{(2)}$ |
| $E_{5\to4}^{\text{mig}} = e_{\text{Al}}^{\text{sp}} - 8\varepsilon_{\text{AlAl}}^{(1)} - 2\varepsilon_{\text{AlX}}^{(1)} - \varepsilon_{\text{AlY}}^{(1)} - 11\varepsilon_{\text{AlV}}^{(1)} - \varepsilon_{\text{XV}}^{(1)} - 5\varepsilon_{\text{AlAl}}^{(2)} - \varepsilon_{\text{AlX}}^{(2)} - 5\varepsilon_{\text{XV}}^{(2)} - \varepsilon_{\text{YV}}^{(2)}$ |
| $E_{6\to5}^{\text{mig}} = e_{\text{X}}^{\text{sp}} - 11\varepsilon_{\text{AlX}}^{(1)} - 8\varepsilon_{\text{AlV}}^{(1)} - 3\varepsilon_{\text{XV}}^{(1)} - \varepsilon_{\text{YV}}^{(1)} - 5\varepsilon_{\text{XX}}^{(2)} - \varepsilon_{\text{XY}}^{(2)} - 6\varepsilon_{\text{AlV}}^{(2)}$ |

**Table S5 Direct cycle starting from a vacancy on the majority sub-lattice.** Activation energies of the forward and backward jumps corresponding to the diffusion of a Y impurity in the $L1_2$ structure of the $Al_3X$ compound.

| Forward jumps |
|---|
| $E_{0\to1}^{\text{mig}} = e_{\text{Y}}^{\text{sp}} - 11\varepsilon_{\text{AlY}}^{(1)} - 8\varepsilon_{\text{AlV}}^{(1)} - 3\varepsilon_{\text{XV}}^{(1)} - \varepsilon_{\text{YV}}^{(1)} - 6\varepsilon_{\text{XY}}^{(2)} - 6\varepsilon_{\text{AlV}}^{(2)}$ |
| $E_{1\to2}^{\text{mig}} = e_{\text{Al}}^{\text{sp}} - 7\varepsilon_{\text{AlAl}}^{(1)} - 3\varepsilon_{\text{AlX}}^{(1)} - \varepsilon_{\text{AlY}}^{(1)} - 11\varepsilon_{\text{AlV}}^{(1)} - \varepsilon_{\text{YV}}^{(1)} - 6\varepsilon_{\text{AlAl}}^{(2)} - 6\varepsilon_{\text{XV}}^{(2)}$ |
| $E_{2\to3}^{\text{mig}} = e_{\text{X}}^{\text{sp}} - 10\varepsilon_{\text{AlX}}^{(1)} - \varepsilon_{\text{XY}}^{(1)} - 8\varepsilon_{\text{AlV}}^{(1)} - 3\varepsilon_{\text{XV}}^{(1)} - \varepsilon_{\text{YV}}^{(1)} - \varepsilon_{\text{AlX}}^{(2)} - 5\varepsilon_{\text{XX}}^{(2)} - 6\varepsilon_{\text{AlV}}^{(2)}$ |
| $E_{3\to4}^{\text{mig}} = e_{\text{Y}}^{\text{sp}} - 8\varepsilon_{\text{AlY}}^{(1)} - 3\varepsilon_{\text{XY}}^{(1)} - 10\varepsilon_{\text{AlV}}^{(1)} - \varepsilon_{\text{XV}}^{(1)} - \varepsilon_{\text{YV}}^{(1)} - 6\varepsilon_{\text{AlY}}^{(2)} - \varepsilon_{\text{AlV}}^{(2)} - 5\varepsilon_{\text{XV}}^{(2)}$ |
| $E_{4\to5}^{\text{mig}} = e_{\text{Al}}^{\text{sp}} - 10\varepsilon_{\text{AlAl}}^{(1)} - \varepsilon_{\text{AlX}}^{(1)} - 8\varepsilon_{\text{AlV}}^{(1)} - 3\varepsilon_{\text{XV}}^{(1)} - \varepsilon_{\text{YV}}^{(1)} - 5\varepsilon_{\text{AlX}}^{(2)} - \varepsilon_{\text{AlY}}^{(2)} - 6\varepsilon_{\text{AlV}}^{(2)}$ |
| $E_{5\to6}^{\text{mig}} = e_{\text{X}}^{\text{sp}} - 8\varepsilon_{\text{AlX}}^{(1)} - 2\varepsilon_{\text{XX}}^{(1)} - \varepsilon_{\text{XY}}^{(1)} - 11\varepsilon_{\text{AlV}}^{(1)} - \varepsilon_{\text{XV}}^{(1)} - 6\varepsilon_{\text{AlX}}^{(2)} - 5\varepsilon_{\text{XV}}^{(2)} - \varepsilon_{\text{YV}}^{(2)}$ |
| **Backward jumps** |
| $E_{1\to0}^{\text{mig}} = e_{\text{Y}}^{\text{sp}} - 8\varepsilon_{\text{AlY}}^{(1)} - 3\varepsilon_{\text{XY}}^{(1)} - 11\varepsilon_{\text{AlV}}^{(1)} - \varepsilon_{\text{YV}}^{(1)} - 6\varepsilon_{\text{AlY}}^{(2)} - 6\varepsilon_{\text{XV}}^{(2)}$ |
| $E_{2\to1}^{\text{mig}} = e_{\text{Al}}^{\text{sp}} - 10\varepsilon_{\text{AlAl}}^{(1)} - \varepsilon_{\text{AlY}}^{(1)} - 8\varepsilon_{\text{AlV}}^{(1)} - 3\varepsilon_{\text{XV}}^{(1)} - \varepsilon_{\text{YV}}^{(1)} - 6\varepsilon_{\text{AlX}}^{(2)} - 6\varepsilon_{\text{AlV}}^{(2)}$ |
| $E_{3\to2}^{\text{mig}} = e_{\text{X}}^{\text{sp}} - 8\varepsilon_{\text{AlX}}^{(1)} - 2\varepsilon_{\text{XX}}^{(1)} - \varepsilon_{\text{XY}}^{(1)} - 10\varepsilon_{\text{AlV}}^{(1)} - \varepsilon_{\text{XV}}^{(1)} - \varepsilon_{\text{YV}}^{(1)} - 6\varepsilon_{\text{AlX}}^{(2)} - \varepsilon_{\text{AlV}}^{(2)} - 5\varepsilon_{\text{XV}}^{(2)}$ |
| $E_{4\to3}^{\text{mig}} = e_{\text{Y}}^{\text{sp}} - 10\varepsilon_{\text{AlY}}^{(1)} - \varepsilon_{\text{XY}}^{(1)} - 8\varepsilon_{\text{AlV}}^{(1)} - 3\varepsilon_{\text{XV}}^{(1)} - \varepsilon_{\text{YV}}^{(1)} - \varepsilon_{\text{AlY}}^{(2)} - 5\varepsilon_{\text{XY}}^{(2)} - 6\varepsilon_{\text{AlV}}^{(2)}$ |
| $E_{5\to4}^{\text{mig}} = e_{\text{Al}}^{\text{sp}} - 7\varepsilon_{\text{AlAl}}^{(1)} - 3\varepsilon_{\text{AlX}}^{(1)} - \varepsilon_{\text{AlY}}^{(1)} - 11\varepsilon_{\text{AlV}}^{(1)} - \varepsilon_{\text{XV}}^{(1)} - 6\varepsilon_{\text{AlAl}}^{(2)} - 5\varepsilon_{\text{XV}}^{(2)} - \varepsilon_{\text{YV}}^{(2)}$ |
| $E_{6\to5}^{\text{mig}} = e_{\text{X}}^{\text{sp}} - 11\varepsilon_{\text{AlX}}^{(1)} - 8\varepsilon_{\text{AlV}}^{(1)} - 3\varepsilon_{\text{XV}}^{(1)} - \varepsilon_{\text{YV}}^{(1)} - 5\varepsilon_{\text{XX}}^{(2)} - \varepsilon_{\text{XY}}^{(2)} - 6\varepsilon_{\text{AlV}}^{(2)}$ |

**Table S6 Bent cycle starting from a vacancy on the majority sub-lattice.** Activation energies of the forward and backward jumps corresponding to the diffusion of a Y impurity in the $L1_2$ structure of the $Al_3X$ compound.

| Forward jumps |
|---|
| $E^{\mathrm{mig}}_{0\to1} = e^{\mathrm{sp}}_{\mathrm{Al}} - 8\varepsilon^{(1)}_{\mathrm{AlAl}} - 2\varepsilon^{(1)}_{\mathrm{AlX}} - \varepsilon^{(1)}_{\mathrm{AlY}} - 12\varepsilon^{(1)}_{\mathrm{AlV}} - 6\varepsilon^{(2)}_{\mathrm{AlAl}} - 5\varepsilon^{(2)}_{\mathrm{XV}} - \varepsilon^{(2)}_{\mathrm{YV}}$ |
| $E^{\mathrm{mig}}_{1\to2} = e^{\mathrm{sp}}_{\mathrm{Y}} - 11\varepsilon^{(1)}_{\mathrm{AlY}} - 9\varepsilon^{(1)}_{\mathrm{AlV}} - 2\varepsilon^{(1)}_{\mathrm{XV}} - \varepsilon^{(1)}_{\mathrm{YV}} - \varepsilon^{(2)}_{\mathrm{AlY}} - 5\varepsilon^{(2)}_{\mathrm{XY}} - 6\varepsilon^{(2)}_{\mathrm{AlV}}$ |
| $E^{\mathrm{mig}}_{2\to3} = e^{\mathrm{sp}}_{\mathrm{Al}} - 9\varepsilon^{(1)}_{\mathrm{AlAl}} - 2\varepsilon^{(1)}_{\mathrm{AlX}} - 11\varepsilon^{(1)}_{\mathrm{AlV}} - \varepsilon^{(1)}_{\mathrm{YV}} - 5\varepsilon^{(2)}_{\mathrm{AlAl}} - \varepsilon^{(2)}_{\mathrm{AlY}} - \varepsilon^{(2)}_{\mathrm{AlV}} - 5\varepsilon^{(2)}_{\mathrm{XV}}$ |
| $E^{\mathrm{mig}}_{3\to4} = e^{\mathrm{sp}}_{\mathrm{Al}} - 10\varepsilon^{(1)}_{\mathrm{AlAl}} - \varepsilon^{(1)}_{\mathrm{AlY}} - 10\varepsilon^{(1)}_{\mathrm{AlV}} - 2\varepsilon^{(1)}_{\mathrm{XV}} - \varepsilon^{(2)}_{\mathrm{AlAl}} - 5\varepsilon^{(2)}_{\mathrm{AlX}} - 5\varepsilon^{(2)}_{\mathrm{AlV}} - \varepsilon^{(2)}_{\mathrm{YV}}$ |
| $E^{\mathrm{mig}}_{4\to5} = e^{\mathrm{sp}}_{\mathrm{Y}} - 9\varepsilon^{(1)}_{\mathrm{AlY}} - 2\varepsilon^{(1)}_{\mathrm{XY}} - 11\varepsilon^{(1)}_{\mathrm{AlV}} - \varepsilon^{(1)}_{\mathrm{YV}} - 6\varepsilon^{(2)}_{\mathrm{AlY}} - \varepsilon^{(2)}_{\mathrm{AlV}} - 5\varepsilon^{(2)}_{\mathrm{XV}}$ |
| $E^{\mathrm{mig}}_{5\to6} = e^{\mathrm{sp}}_{\mathrm{Al}} - 11\varepsilon^{(1)}_{\mathrm{AlAl}} - 9\varepsilon^{(1)}_{\mathrm{AlV}} - 2\varepsilon^{(1)}_{\mathrm{XV}} - \varepsilon^{(1)}_{\mathrm{YV}} - 5\varepsilon^{(2)}_{\mathrm{AlX}} - \varepsilon^{(2)}_{\mathrm{AlY}} - 6\varepsilon^{(2)}_{\mathrm{AlV}}$ |
| **Backward jumps** |
| $E^{\mathrm{mig}}_{1\to0} = E^{\mathrm{mig}}_{5\to6}$ |
| $E^{\mathrm{mig}}_{2\to1} = E^{\mathrm{mig}}_{4\to5}$ |
| $E^{\mathrm{mig}}_{3\to2} = E^{\mathrm{mig}}_{3\to4}$ |
| $E^{\mathrm{mig}}_{4\to3} = E^{\mathrm{mig}}_{2\to3}$ |
| $E^{\mathrm{mig}}_{5\to4} = E^{\mathrm{mig}}_{1\to2}$ |
| $E^{\mathrm{mig}}_{6\to5} = E^{\mathrm{mig}}_{0\to1}$ |

Table S7 **Direct cycle starting from a vacancy on the minority sub-lattice.** Activation energies of the forward and backward jumps corresponding to the diffusion of a Y impurity in the $L1_2$ structure of the $Al_3X$ compound.

| Forward jumps |
|---|
| $E^{\mathrm{mig}}_{0\to1} = e^{\mathrm{sp}}_{\mathrm{Al}} - 8\varepsilon^{(1)}_{\mathrm{AlAl}} - 2\varepsilon^{(1)}_{\mathrm{AlX}} - \varepsilon^{(1)}_{\mathrm{AlY}} - 12\varepsilon^{(1)}_{\mathrm{AlV}} - 6\varepsilon^{(2)}_{\mathrm{AlAl}} - 5\varepsilon^{(2)}_{\mathrm{XV}} - \varepsilon^{(2)}_{\mathrm{YV}}$ |
| $E^{\mathrm{mig}}_{1\to2} = e^{\mathrm{sp}}_{\mathrm{Y}} - 11\varepsilon^{(1)}_{\mathrm{AlY}} - 9\varepsilon^{(1)}_{\mathrm{AlV}} - 2\varepsilon^{(1)}_{\mathrm{XV}} - \varepsilon^{(1)}_{\mathrm{YV}} - \varepsilon^{(2)}_{\mathrm{AlY}} - 5\varepsilon^{(2)}_{\mathrm{XY}} - 6\varepsilon^{(2)}_{\mathrm{AlV}}$ |
| $E^{\mathrm{mig}}_{2\to3} = e^{\mathrm{sp}}_{\mathrm{Al}} - 8\varepsilon^{(1)}_{\mathrm{AlAl}} - 2\varepsilon^{(1)}_{\mathrm{AlX}} - \varepsilon^{(1)}_{\mathrm{AlY}} - 11\varepsilon^{(1)}_{\mathrm{AlV}} - \varepsilon^{(1)}_{\mathrm{YV}} - 6\varepsilon^{(2)}_{\mathrm{AlAl}} - \varepsilon^{(2)}_{\mathrm{AlV}} - 5\varepsilon^{(2)}_{\mathrm{XV}}$ |
| $E^{\mathrm{mig}}_{3\to4} = e^{\mathrm{sp}}_{\mathrm{Al}} - 10\varepsilon^{(1)}_{\mathrm{AlAl}} - \varepsilon^{(1)}_{\mathrm{AlY}} - 9\varepsilon^{(1)}_{\mathrm{AlV}} - 2\varepsilon^{(1)}_{\mathrm{XV}} - \varepsilon^{(1)}_{\mathrm{YV}} - \varepsilon^{(2)}_{\mathrm{AlAl}} - 5\varepsilon^{(2)}_{\mathrm{AlX}} - 6\varepsilon^{(2)}_{\mathrm{AlV}}$ |
| $E^{\mathrm{mig}}_{4\to5} = e^{\mathrm{sp}}_{\mathrm{Y}} - 9\varepsilon^{(1)}_{\mathrm{AlY}} - 2\varepsilon^{(1)}_{\mathrm{XY}} - 11\varepsilon^{(1)}_{\mathrm{AlV}} - \varepsilon^{(1)}_{\mathrm{YV}} - 6\varepsilon^{(2)}_{\mathrm{AlY}} - \varepsilon^{(2)}_{\mathrm{AlV}} - 5\varepsilon^{(2)}_{\mathrm{XV}}$ |
| $E^{\mathrm{mig}}_{5\to6} = e^{\mathrm{sp}}_{\mathrm{Al}} - 11\varepsilon^{(1)}_{\mathrm{AlAl}} - 9\varepsilon^{(1)}_{\mathrm{AlV}} - 2\varepsilon^{(1)}_{\mathrm{XV}} - \varepsilon^{(1)}_{\mathrm{YV}} - 5\varepsilon^{(2)}_{\mathrm{AlX}} - \varepsilon^{(2)}_{\mathrm{AlY}} - 6\varepsilon^{(2)}_{\mathrm{AlV}}$ |
| **Backward jumps** |
| $E^{\mathrm{mig}}_{1\to0} = E^{\mathrm{mig}}_{5\to6}$ |
| $E^{\mathrm{mig}}_{2\to1} = E^{\mathrm{mig}}_{4\to5}$ |
| $E^{\mathrm{mig}}_{3\to2} = E^{\mathrm{mig}}_{3\to4}$ |
| $E^{\mathrm{mig}}_{4\to3} = E^{\mathrm{mig}}_{2\to3}$ |
| $E^{\mathrm{mig}}_{5\to4} = E^{\mathrm{mig}}_{1\to2}$ |
| $E^{\mathrm{mig}}_{6\to5} = E^{\mathrm{mig}}_{0\to1}$ |

Table S8 **Bent cycle starting from a vacancy on the minority sub-lattice.** Activation energies of the forward and backward jumps corresponding to the diffusion of a Y impurity in the $L1_2$ structure of the $Al_3X$ compound.